\documentclass{article}

\usepackage{arxiv}

\usepackage[utf8]{inputenc} % allow utf-8 input
\usepackage[T1]{fontenc}    % use 8-bit T1 fonts
\usepackage{hyperref}       % hyperlinks
\usepackage{url}            % simple URL typesetting
\usepackage{booktabs}       % professional-quality tables
\usepackage{amsfonts}       % blackboard math symbols
\usepackage{nicefrac}       % compact symbols for 1/2, etc.
\usepackage{microtype}      % microtypography
\usepackage{lipsum}
\usepackage{graphicx}
\usepackage{tabularx}
\usepackage{amsmath}
\usepackage{algorithm}
\usepackage{algpseudocode}
\graphicspath{ {./images/} }

\title{You Only Convert Twice: Ozaki Scheme II for Chained Tensor Mode Products}

\author{
 Xuanzhengbo Ren \\
  Information Technology Center\\
  The University of Tokyo\\
  Kashiwanoha, Chiba, 2770882 Japan \\
  \texttt{xuanzhengbo.ren@cc.u-tokyo.ac.jp} \\
}

\begin{document}
\maketitle
%% All algorithm definitions, loaded once by main.tex.
% Section files invoke these commands where the algorithms should appear.
% Requires algorithm, algpseudocode, amsmath, and amssymb.

\newcommand{\RnsDropAlgorithm}{%
\begin{algorithm}[t]
\caption{RnsDrop for rescaling by one modulus}
\label{alg:rnsdrop}
\small
\begin{algorithmic}[1]
\Require Centered residue planes $R$ representing integer entries $v$
         on $\mathcal{L}$, drop modulus
         $q\in\mathcal{L}$, scale $\alpha$
\Require The planned quotient satisfies $2B'<P_{\mathcal{L}}/q$
\Ensure Residues of $v'=(v-[v]_q)/q$ on $\mathcal{L}\setminus\{q\}$,
        with decoding scale $\alpha'=\alpha q$
\Procedure{RnsDrop}{$R,\mathcal{L},q,\alpha$}
\State $S\gets R_q$
\ForAll{$p\in\mathcal{L}\setminus\{q\}$}
  \State $R_p\gets[(R_p-S)(q^{-1}\bmod p)]_p$
\EndFor
\State Remove plane $R_q$ and modulus $q$ from $\mathcal{L}$
\State Record $q$ in the scale metadata for $\alpha' = \alpha q$
\State \Return $(R,\mathcal{L},\alpha')$
\EndProcedure
\end{algorithmic}
\end{algorithm}
}

\newcommand{\DropPlanAlgorithm}{%
\begin{algorithm}[t]
\caption{DropPlan from operator row-sum bounds, independent of the input tensor}
\label{alg:dropplan}
\small
\begin{algorithmic}[1]
\Require Encoding parameter $b$, row-sum bounds $(\rho_i)_{i=1}^d$,
         basis $\mathcal{P}$, restoration flag $\tau\in\{0,1\}$
\Ensure Drop counts $(g_i)_{i=1}^d$ or failure $\bot$
\Procedure{DropPlan}{$b,(\rho_i),\mathcal{P},\tau$}
\State $B\gets2^b$
\State $\mathcal{L}\gets\mathcal{P}$
\State $P_{\mathcal{L}}\gets\prod_{p\in\mathcal{L}}p$
\For{$i=1,\ldots,d$}
  \State $B\gets B\rho_i$
  \If{$2B\geq P_{\mathcal{L}}$}
    \State \Return $\bot$
  \EndIf
  \State $g_i\gets0$
  \If{$i<d$}
    \While{$|\mathcal{L}|>1$ and $\lfloor B/\min\mathcal{L}\rfloor\geq2^b$}
      \State $q\gets\min\mathcal{L}$
      \State $B'\gets\lfloor B/q\rfloor+1$
      \If{$2B'\geq P_{\mathcal{L}}/q$}
        \State \Return $\bot$
      \EndIf
      \State $(\mathcal{L},P_{\mathcal{L}},B)\gets
             (\mathcal{L}\setminus\{q\},P_{\mathcal{L}}/q,B')$
      \State $g_i\gets g_i+1$
    \EndWhile
    \If{$\tau=1$}
      \State $\mathcal{L}\gets\mathcal{P}$
      \State $P_{\mathcal{L}}\gets\prod_{p\in\mathcal{P}}p$
    \EndIf
  \EndIf
\EndFor
\State \Return $(g_1,\ldots,g_d)$
\EndProcedure
\end{algorithmic}
\end{algorithm}
}

\newcommand{\PlanRescaleBEAlgorithm}{%
\begin{algorithm}[t]
\caption{Planner for precision selection on a fixed working basis}
\label{alg:basisplan}
\small
\begin{algorithmic}[1]
\Require Operator row-sum bound functions $(\rho_i)$,
         fixed working basis $\mathcal{P}$ with $s=|\mathcal{P}|$
\Ensure Selected encoding parameter $b$ and drop counts $g$, or failure $\bot$
\Procedure{Planner}{$(\rho_i),\mathcal{P}$}
  \State $b\gets53$
  \State $g\gets\Call{DropPlan}{b,(\rho_i(b)),\mathcal{P},1}$
  \While{$g=\bot$}
    \State $b\gets b-1$
    \If{$b<1$}
      \State \Return $\bot$
    \EndIf
    \State $g\gets\Call{DropPlan}{b,(\rho_i(b)),\mathcal{P},1}$
  \EndWhile
  \State \Return $(b,g)$
\EndProcedure
\end{algorithmic}
\end{algorithm}
}

\newcommand{\BaseExtensionAlgorithm}{%
\begin{algorithm}[t]
\caption{MrcDigits and BaseExtend for centered residue conversion}
\label{alg:mrc}
\small
\begin{algorithmic}[1]
\Require Centered residues $R$ of $v'$ on reduced basis
         $\mathcal{P}'=\{q_1,\ldots,q_m\}$, processed in that order
\Require $2|v'|<P'$, where $P'=\prod_iq_i$,
         with the even modulus first if present
\Require Removed moduli $\mathcal{D}=\mathcal{P}\setminus\mathcal{P}'$
\Require Precomputed inverses and weights
         $\omega_{i,p}=[\prod_{j<i}q_j]_p$
\Ensure Base extension represents the same integer $v'$ on
        $\mathcal{P}'\cup\mathcal{D}=\mathcal{P}$
\Procedure{MrcDigits}{$R,\mathcal{P}'$}
  \State $U\gets R$ \Comment{working copy of the input residues}
  \For{$i=1,\ldots,m$}
    \State $d_i\gets U_i$
    \For{$j=i+1,\ldots,m$}
      \State $U_j\gets[(U_j-d_i)(q_i^{-1}\bmod q_j)]_{q_j}$
    \EndFor
  \EndFor
  \State \Return $(d_1,\ldots,d_m)$
\EndProcedure
\Statex
\Procedure{BaseExtend}{$R,\mathcal{P}',\mathcal{D}$}
\State $(d_1,\ldots,d_m)\gets\Call{MrcDigits}{R,\mathcal{P}'}$
\ForAll{$p\in\mathcal{D}$}
  \State $t\gets0$
  \For{$i=1,\ldots,m$}
    \State $t\gets[t+d_i\omega_{i,p}]_p$
  \EndFor
  \State $R_p\gets t$
\EndFor
\State \Return $R$
\EndProcedure
\end{algorithmic}
\end{algorithm}
}

\newcommand{\TensorEncodingAlgorithm}{%
\begin{algorithm}[t]
\caption{Encoding and reconstruction primitives}
\label{alg:encode}
\small
\begin{algorithmic}[1]
\Procedure{Encode}{$T,b,\mathcal{G}$}
  \ForAll{groups $G\in\mathcal{G}$}
    \If{$\max|G|=0$}
      \State $(\bar G,e_G)\gets(0,0)$
    \Else
      \State Choose $E_G$ with $2^{E_G-1}\leq\max|G|<2^{E_G}$
      \State $e_G\gets b-E_G$
      \State $\bar G\gets\operatorname{round}(2^{e_G}G)$
    \EndIf
  \EndFor
  \State \Return $(\bar T,e)$
\EndProcedure
\Statex
\Procedure{ToRns}{$\bar T,\mathcal{L}$}
  \State \Return $\{R_p=[\bar T]_p\}_{p\in\mathcal{L}}$
\EndProcedure
\Statex
\Procedure{ReconstructFP64}{$R,\mathcal{L},\sigma$}
  \State Order $\mathcal{L}=\{q_1,\ldots,q_m\}$ with the even modulus first
  \State $(d_1,\ldots,d_m)\gets\Call{MrcDigits}{R,\mathcal{L}}$
  \State $\widehat V\gets\operatorname{fl}_{64}(d_m)$
  \For{$j=m-1,\ldots,1$}
    \State $\widehat V\gets\operatorname{fl}_{64}(q_j\widehat V+d_j)$
  \EndFor
  \State \Return $\operatorname{fl}_{64}(\sigma\odot\widehat V)$
\EndProcedure
\end{algorithmic}
\end{algorithm}
}

\newcommand{\TensorComposedAlgorithm}{%
\begin{algorithm}[t]
\caption{OzII-Composed for successive tensor mode products}
\label{alg:OzIIcomposed}
\small
\begin{algorithmic}[1]
\Require Tensor $X$, operators $(A_i)_{i=1}^d$, modes $(m_i)_{i=1}^d$
\Require Fixed residue basis $\mathcal{P}$ with $s$ pairwise coprime moduli
\Procedure{OzIIComposed}{$X,\mathcal{P},(A_i),(m_i)$}
  \State $P\gets\prod_{p\in\mathcal{P}}p$
  \State Choose the largest $b\in\{1,\ldots,53\}$ with $n2^{2b}<P/2$
  \If{no such $b$ exists}
    \State \Return $\bot$
  \EndIf
  \For{$i=1,\ldots,d$}
    \State $(\bar A_i,f_i)\gets\Call{Encode}{A_i,b,\operatorname{Rows}(A_i)}$
    \State Cache $\bar A_{i,p}=[\bar A_i]_p$ for $p\in\mathcal{P}$
  \EndFor
  \State $T\gets X$
  \For{$i=1,\ldots,d$}
    \State $(\bar T,e_i)\gets\Call{Encode}{T,b,\operatorname{Fibers}_{m_i}(T)}$
    \State $R\gets\Call{ToRns}{\bar T,\mathcal{P}}$
    \ForAll{$p\in\mathcal{P}}$
      \State $R_p\gets[R_p\times_{m_i}\bar A_{i,p}]_p$
    \EndFor
    \State $\sigma_i\gets2^{-e_i}\,2^{-f_i}$ with output-index broadcasting
    \State $T\gets\Call{ReconstructFP64}{R,\mathcal{P},\sigma_i}$
  \EndFor
  \State \Return $T$
\EndProcedure
\end{algorithmic}
\end{algorithm}
}

\newcommand{\TensorPersistentAlgorithm}{%
\begin{algorithm}[t]
\caption{OzII-RescaleBE for persistent tensor mode products}
\label{alg:OzIIrescalebe}
\small
\begin{algorithmic}[1]
\Require Tensor $X$, operators $(A_i)_{i=1}^d$, modes $(m_i)_{i=1}^d$
\Require Fixed residue basis $\mathcal{P}$ with $s$ pairwise coprime moduli
\Procedure{OzIIRescaleBE}{$X,\mathcal{P},(A_i),(m_i)$}
  \State Define $(\rho_i(b))$ from effective operator encoding at precision $b$
  \State $\pi\gets\Call{PlanRescaleBE}{(\rho_i),\mathcal{P}}$
  \If{$\pi=\bot$}
    \State \Return $\bot$
  \EndIf
  \State $(b,(g_i))\gets\pi$
  \State Prepare effective operators $(\bar A_i)$ and scale field $\Gamma$ at $b$
  \State Cache $\bar A_{i,p}=[\bar A_i]_p$ for $p\in\mathcal{P}$
  \State $(\bar X,e)\gets\Call{Encode}{X,b,\{X\}}$
  \State $\alpha\gets2^{-e}$
  \State $R\gets\Call{ToRns}{\bar X,\mathcal{P}}$
  \For{$i=1,\ldots,d$}
    \ForAll{$p\in\mathcal{P}$}
      \State $R_p\gets[R_p\times_{m_i}\bar A_{i,p}]_p$
    \EndFor
    \If{$g_i>0$}
      \State $\mathcal{L}\gets\mathcal{P}$
      \For{$j=1,\ldots,g_i$}
        \State $q\gets\min\mathcal{L}$
        \State $(R,\mathcal{L},\alpha')\gets\Call{RnsDrop}{R,\mathcal{L},q,\alpha}$
        \State $\alpha\gets\alpha'$
      \EndFor
      \State $\mathcal{P}'\gets\mathcal{L}$
      \State $\mathcal{D}\gets\mathcal{P}\setminus\mathcal{P}'$
      \State $R\gets\Call{BaseExtend}{R,\mathcal{P}',\mathcal{D}}$
      \State $\mathcal{L}\gets\mathcal{P}$
    \EndIf
  \EndFor
  \State \Return $\Call{ReconstructFP64}{R,\mathcal{P},\alpha\Gamma}$
\EndProcedure
\end{algorithmic}
\end{algorithm}
}

%%%%%%%%%%%%%%%%%%%%%%%%%%%%5 For Appendix %%%%%%%%%%%%%%%%%%%%%%%%%%%%%%%%%%%%%
\newcommand{\PlanRescaleAppendixAlgorithm}{%
\begin{algorithm}[t]
\caption{PlanRescale for precision selection without base extension}
\label{alg:app-planrescale}
\small
\begin{algorithmic}[1]
\Require Operator row-sum bound functions $(\rho_i)$, initial basis $\mathcal P$
\Ensure Encoding parameter $b$ and drop counts $g$, or failure $\bot$
\Procedure{PlanRescale}{$(\rho_i),\mathcal P$}
  \For{$b=53,52,\ldots,1$}
    \State $g\gets\Call{DropPlan}{b,(\rho_i(b)),\mathcal P,0}$
    \If{$g\neq\bot$}
      \State \Return $(b,g)$
    \EndIf
  \EndFor
  \State \Return $\bot$
\EndProcedure
\end{algorithmic}
\end{algorithm}
}

\newcommand{\NoRescaleAppendixAlgorithm}{%
\begin{algorithm}[t]
\caption{OzII-NoRescale with a fixed basis throughout the chain}
\label{alg:app-OzIInorescale}
\small
\begin{algorithmic}[1]
\Require Tensor $X$, operators $(A_i)_{i=1}^{d}$, modes $(m_i)_{i=1}^{d}$
\Require Initial basis $\mathcal P$ with $s=|\mathcal P|$
\Procedure{OzIINoRescale}{$X,\mathcal P,(A_i),(m_i)$}
  \State $P\gets\prod_{p\in\mathcal P}p$
  \State Define $(\rho_i(b))$ from effective operator encoding
  \State Choose the largest $b\in\{1,\ldots,53\}$ with $2B_j(b)<P$ for all $j=0,\ldots,d$
  \If{no such $b$ exists}
    \State \Return $\bot$
  \EndIf
  \State Prepare effective operators $(\bar A_i)$ and scale field $\Gamma$ at $b$
  \State Cache $\bar A_{i,p}=[\bar A_i]_p$ for $p\in\mathcal P$
  \State $(\bar X,e)\gets\Call{Encode}{X,b,\{X\}}$
  \State $\alpha\gets2^{-e}$
  \State $R\gets\Call{ToRns}{\bar X,\mathcal P}$
  \For{$i=1,\ldots,d$}
    \ForAll{$p\in\mathcal P$}
      \State $R_p\gets[R_p\times_{m_i}\bar A_{i,p}]_p$
    \EndFor
  \EndFor
  \State \Return $\Call{ReconstructFP64}{R,\mathcal P,\alpha\Gamma}$
\EndProcedure
\end{algorithmic}
\end{algorithm}
}

\newcommand{\RescaleAppendixAlgorithm}{%
\begin{algorithm}[t]
\caption{OzII-Rescale with a shrinking residue basis}
\label{alg:app-OzIIrescale}
\small
\begin{algorithmic}[1]
\Require Tensor $X$, operators $(A_i)_{i=1}^{d}$, modes $(m_i)_{i=1}^{d}$
\Require Initial basis $\mathcal P$ with $s=|\mathcal P|$
\Procedure{OzIIRescale}{$X,\mathcal P,(A_i),(m_i)$}
  \State Define $(\rho_i(b))$ from effective operator encoding
  \State $\pi\gets\Call{PlanRescale}{(\rho_i),\mathcal P}$
  \If{$\pi=\bot$}
    \State \Return $\bot$
  \EndIf
  \State $(b,(g_i))\gets\pi$
  \State Prepare effective operators $(\bar A_i)$ and scale field $\Gamma$ at $b$
  \State Cache $\bar A_{i,p}=[\bar A_i]_p$ for $p\in\mathcal P$
  \State $(\bar X,e)\gets\Call{Encode}{X,b,\{X\}}$
  \State $\alpha\gets2^{-e}$
  \State $\mathcal L\gets\mathcal P$
  \State $R\gets\Call{ToRns}{\bar X,\mathcal L}$
  \For{$i=1,\ldots,d$}
    \ForAll{$p\in\mathcal L$}
      \State $R_p\gets[R_p\times_{m_i}\bar A_{i,p}]_p$
    \EndFor
    \For{$j=1,\ldots,g_i$}
      \State $q\gets\min\mathcal L$
      \State $(R,\mathcal L,\alpha')\gets\Call{RnsDrop}{R,\mathcal L,q,\alpha}$
      \State $\alpha\gets\alpha'$
    \EndFor
  \EndFor
  \State \Return $\Call{ReconstructFP64}{R,\mathcal L,\alpha\Gamma}$
\EndProcedure
\end{algorithmic}
\end{algorithm}
}
% All manuscript table definitions, loaded once by main.tex.
% Requires booktabs, tabularx, and array.

\newcommand{\ExperimentalEnvironmentTable}{%
\begin{table}[t]
  \centering
  \caption{Hardware and software environment.}
  \label{tab:experimental-environment}
  \small
  \renewcommand{\arraystretch}{1.12}
  \begin{tabularx}{\linewidth}{@{}>{\raggedright\arraybackslash}p{0.43\linewidth}
                                   >{\raggedright\arraybackslash}X@{}}
    \toprule
    \multicolumn{2}{@{}l}{\textbf{Hardware}} \\
    \midrule
    GPU & NVIDIA RTX PRO 6000 Blackwell Max-Q Workstation Edition \\
    CUDA target & \texttt{sm\_120} \\
    Streaming multiprocessors & 188 \\
    Device memory & 96 GB GDDR7 \\
    Peak memory bandwidth & 1792 GB/s \\
    Peak FP64 throughput & $\approx 1.71$ TFLOP/s \\
    Peak INT8 Tensor Core throughput & 877.9 TOP/s \\
    \midrule
    \multicolumn{2}{@{}l}{\textbf{Software}} \\
    \midrule
    CuTe DSL & 4.5.2 \\
    PyTorch & 2.11.0 \\
    CUDA & 13.3 \\
    cuBLAS & 13.6 \\
    \bottomrule
  \end{tabularx}
\end{table}
}

\newcommand{\ElementwiseAccuracyTable}{%
\begin{table*}[t]
  \centering
  \small
  \setlength{\tabcolsep}{3pt}
  \renewcommand{\arraystretch}{1.12}
  \caption{Maximum and median elementwise errors at depths 3 and 8,
  using 32 tensors per dataset with $n=8$.
  All residue methods start with $s=15$. }
  \label{tab:accuracy-elementwise}
  \begin{tabular}{@{}lc*{6}{l}@{}}
    \toprule
    Method & $d$ & \multicolumn{2}{c}{Uniform}
                & \multicolumn{2}{c}{Smooth}
                & \multicolumn{2}{c}{Wide160} \\
    \cmidrule(lr){3-4}\cmidrule(lr){5-6}\cmidrule(l){7-8}
    & & Max $\varepsilon_{\mathrm{max}}$ & Median $\varepsilon_{\mathrm{med}}$ & Max $\varepsilon_{\mathrm{max}}$ & Median $\varepsilon_{\mathrm{med}}$ & Max $\varepsilon_{\mathrm{max}}$ & Median $\varepsilon_{\mathrm{med}}$ \\
    \midrule
    FP64 einsum & 3 & $8.26\times10^{-12}$ & $1.93\times10^{-16}$ & $9.35\times10^{-11}$ & $4.33\times10^{-16}$ & $1.09\times10^{-12}$ & $1.61\times10^{-16}$ \\
    OzII-Composed & 3 & $3.95\times10^{-12}$ & $2.53\times10^{-16}$ & $8.47\times10^{-10}$ & $6.53\times10^{-16}$ & $9.18\times10^{-13}$ & $3.05\times10^{-16}$ \\
    OzII-NoRescale & 3 & $1.55\times10^{-5}$ & $7.82\times10^{-9}$ & $1.44\times10^{-1}$ & $4.00\times10^{-8}$ & $1.44\times10^{-3}$ & $1.72\times10^{-8}$ \\
    OzII-Rescale & 3 & $4.58\times10^{-5}$ & $8.03\times10^{-9}$ & $1.72\times10^{-1}$ & $4.26\times10^{-8}$ & $1.99\times10^{-3}$ & $2.54\times10^{-8}$ \\
    OzII-RescaleBE & 3 & $9.49\times10^{-13}$ & $1.87\times10^{-16}$ & $4.52\times10^{-9}$ & $1.47\times10^{-15}$ & $3.90\times10^{-11}$ & $1.54\times10^{-15}$ \\
    GEMMul8-Composed & 3 & $3.09\times10^{-12}$ & $1.51\times10^{-16}$ & $1.24\times10^{-10}$ & $2.09\times10^{-16}$ & $9.17\times10^{-13}$ & $1.64\times10^{-16}$ \\
    GEMMul8-KRON & 3 & $2.37\times10^{-12}$ & $7.28\times10^{-16}$ & $7.54\times10^{-9}$ & $2.67\times10^{-15}$ & $4.55\times10^{-12}$ & $4.49\times10^{-16}$ \\
    \midrule
    FP64 einsum & 8 & $2.56\times10^{-12}$ & $3.82\times10^{-16}$ & $4.67\times10^{-12}$ & $8.56\times10^{-16}$ & $5.79\times10^{-11}$ & $3.78\times10^{-16}$ \\
    OzII-Composed & 8 & $1.29\times10^{-11}$ & $5.39\times10^{-16}$ & $5.28\times10^{-12}$ & $1.24\times10^{-15}$ & $2.84\times10^{-11}$ & $6.21\times10^{-16}$ \\
    OzII-NoRescale & 8 & $8.27$ & $1.48\times10^{-3}$ & $7.10\times10^{1}$ & $6.18\times10^{-3}$ & $3.53\times10^{1}$ & $2.27\times10^{-3}$ \\
    OzII-Rescale & 8 & $4.06\times10^{3}$ & $1.00$ & $4.54\times10^{1}$ & $1.00$ & $1.00$ & $1.00$ \\
    OzII-RescaleBE & 8 & $6.79\times10^{-10}$ & $1.56\times10^{-13}$ & $5.35\times10^{-8}$ & $2.00\times10^{-12}$ & $9.96\times10^{-8}$ & $2.27\times10^{-12}$ \\
    GEMMul8-Composed & 8 & $1.13\times10^{-12}$ & $2.67\times10^{-16}$ & $2.45\times10^{-12}$ & $5.58\times10^{-16}$ & $1.13\times10^{-11}$ & $3.15\times10^{-16}$ \\
    GEMMul8-KRON & 8 & $5.27\times10^{-12}$ & $8.04\times10^{-16}$ & $2.39\times10^{-11}$ & $1.81\times10^{-15}$ & $1.86\times10^{-11}$ & $5.23\times10^{-16}$ \\
    \bottomrule
  \end{tabular}
\end{table*}
}

\newcommand{\ModuliPerStageTable}{%
\begin{table*}[t]
  \centering
  \small
  \setlength{\tabcolsep}{5pt}
  \renewcommand{\arraystretch}{1.12}
  \caption{Selected encoding parameter $b$ for the accuracy tests with
  $n=8$ and initial modulus count $s=15$ for every residue method.}
  \label{tab:moduli}
  \begin{tabular}{@{}lcccccc@{}}
    \toprule
    Method & $d=3$ & $d=4$ & $d=5$ & $d=6$ & $d=7$ & $d=8$ \\
    \midrule
    FP64 einsum & --- & --- & --- & --- & --- & --- \\
    OzII-Composed & 53 & 53 & 53 & 53 & 53 & 53 \\
    OzII-NoRescale & 27 & 21 & 17 & 14 & 12 & 10 \\
    OzII-Rescale & 27 & 21 & 17 & 14 & 12 & 10 \\
    OzII-RescaleBE & 53 & 53 & 53 & 53 & 53 & 53 \\
    GEMMul8-Composed & 53 & 53 & 53 & 53 & 53 & 53 \\
    GEMMul8-KRON & 53 & 53 & 53 & 53 & 53 & 53 \\
    \bottomrule
  \end{tabular}
\end{table*}
}

%%%%%%%%%%%%%%%%%%%%%%%%%%%%% Discarded %%%%%%%%%%%%%%%%%%%%%%%%%%%%%%%
% \newcommand{\MemoryTrafficTable}{%
% \begin{table*}[t]
%   \centering
%   \small
%   \setlength{\tabcolsep}{5pt}
%   \renewcommand{\arraystretch}{1.12}
%   \caption{Measured DRAM traffic in bytes per element-local
%   DoF for $2^{24}$ DoFs at $d=3$ and $n=8, 256$, $s=15$.}
%   \label{tab:memory-traffic}
%   \begin{tabular}{@{}lrrrrrr@{}}
%     \toprule
%     & \multicolumn{3}{c}{$n=8$}
%     & \multicolumn{3}{c}{$n=256$} \\
%     \cmidrule(lr){2-4}\cmidrule(lr){5-7}
%     Method & Read & Write & Total & Read & Write & Total \\
%     \midrule
%     FP64 einsum
%       & 48.0 & 27.7 & 75.7
%       & 41.4 & 20.3 & 61.7 \\
%     OzII-Composed
%       & 183.1 & 93.3 & 276.4
%       & 165.0 & 100.1 & 265.1 \\
%     OzII-RescaleBE
%       & 114.2 & 79.8 & 194.0
%       & 116.3 & 82.0 & 198.3 \\
%     GEMMul8-Composed
%       & 7588.1 & 4085.2 & 11673.3
%       & 366.2 & 102.4 & 468.6 \\
%     GEMMul8-Composed (fast)
%       & 7466.4 & 2536.1 & 10002.5
%       & 327.3 & 106.5 & 433.8 \\
%     GEMMul8-KRON
%       & 111.5 & 44.2 & 155.7
%       & --- & --- & --- \\
%     \bottomrule
%   \end{tabular}
% \end{table*}
% }

\begin{abstract}
Ozaki scheme~II enables high-precision matrix multiplication
using low-precision Tensor Cores through an integer residue
representation.
Applying it to successive tensor mode products offers a way
to exploit the low-precision performance of recent GPUs.
This work proposes OzII-RescaleBE, a method that performs
rescaling and base extension directly on residue representations.
% It enables persistent execution of chained tensor mode products,
% keeping intermediate tensors in residue form without converting
% them back to FP64.
It enables persistent execution of chained tensor mode products,
keeping intermediate tensors in residue form.
Conversion to residues is required only at the beginning of
the chain, followed by reconstruction at the end.
An implementation using CuTe DSL and INT8 Tensor Cores is
provided for verification and evaluation.
The method is compared with per-mode composition of Ozaki
scheme~II (GEMMul8-Composed) and a dense Kronecker-product
formulation (GEMMul8-KRON), both implemented using the
GEMMul8 library.
Across three datasets with different input distributions,
OzII-RescaleBE achieves normwise relative errors ranging
from $2.5\times10^{-16}$ to $2.6\times10^{-15}$ for
chain depths $d=3$ through $7$.
The corresponding errors range from $10^{-16}$ to $10^{-15}$
for GEMMul8-KRON and from $1\times10^{-16}$ to
$3\times10^{-16}$ for GEMMul8-Composed and the FP64 chain.
At $d=8$, the errors of OzII-RescaleBE increase to between
$7\times10^{-14}$ and $6\times10^{-13}$.
For mode sizes $n=8$ to $256$, OzII-RescaleBE achieves
throughputs ranging from 1.47 to 4.42\,GDoF/s.
Its throughput is within 12\,\% of GEMMul8-KRON at
$n=8$ and $12$ and exceeds it at larger tested sizes.
Compared with GEMMul8-Composed, OzII-RescaleBE achieves
$2.1$--$60\times$ the throughput for $n\leq128$,
but has 6--12\,\% lower throughput at $n=256$.
It also uses less device memory than the GEMMul8 baselines
in the measured memory comparisons.
These results demonstrate that rescaling and base extension
enable accurate residue-domain tensor chains while avoiding
repeated intermediate conversions to floating point.
\end{abstract}

% keywords can be removed
 
\section{Introduction}
\label{sec:introduction}

The growing demand for artificial intelligence training and inference
is shaping GPU architectures around high-throughput, low-precision
matrix arithmetic.
However, improvements in these capabilities do not necessarily
extend to native double-precision (FP64) computation.
For example, NVIDIA lists 34~TFLOP/s of conventional FP64 throughput
for the H100 SXM, whereas an individual Blackwell Ultra GPU in the
HGX B300 is specified at 1.2~TFLOP/s of FP64 throughput and
2.25~PFLOP/s of dense FP16/BF16 Tensor Core
throughput~\cite{nvidiaH100,nvidiaBlackwellUltra}.
For scientific applications that depend on FP64 accuracy, this disparity motivates algorithms that exploit low-precision matrix units while retaining higher-precision numerical behavior.
Mixed-precision methods exploit faster lower-precision
arithmetic by combining computations at different
precisions, reserving higher precision for operations
that are more sensitive to numerical error~\cite{higham2022mixed}.
Precision emulation provides a complementary approach by constructing higher-precision arithmetic from multiple low-precision operations.

The Ozaki schemes offer a promising foundation for such emulation.
Originally developed using error-free transformations of matrix
products, they have since been adapted to floating-point and
integer Tensor Cores~\cite{ozaki2012eft,mukunoki2020tensor,ootomo2024integer}.
Among these approaches, Ozaki scheme~II uses an integer residue
representation to perform matrix multiplication on low-precision
integer hardware~\cite{ozaki2025scheme2}.
It first scales and truncates the floating-point operands to
integers, then converts them to residues with respect to several
pairwise coprime moduli.
This conversion process is referred to as encoding in this paper.
Matrix products are then computed independently for each modulus,
and the results are reconstructed using the Chinese remainder
theorem (CRT) and scaled back to floating point.
GEMMul8 provides a public implementation of this
approach~\cite{gemmul8}.

We consider its application to chained tensor mode products, a central building block of sum factorization in high-order finite element methods.
For tensor-product elements, sum factorization applies one-dimensional operators successively along the tensor modes, avoiding the formation and application of a full multidimensional matrix~\cite{swirydowicz2019tensor}.
A representative three-dimensional operation is
\begin{equation}
  Y = X \times_1 A \times_2 B \times_3 C,
  \qquad
  X \in \mathbb{R}^{n\times n\times n},
  \quad A,B,C \in \mathbb{R}^{n\times n}.
  \label{eq:intro-mode-chain}
\end{equation}
Here, $\times_j$ denotes multiplication along mode~$j$.
Each mode product can be expressed as a matrix multiplication through an appropriate unfolding. 
For the square case in \eqref{eq:intro-mode-chain}, this reduces the arithmetic cost from $O(n^6)$ for a dense operator application to $O(n^4)$.
The workload therefore appears well suited to GEMM-based emulation, but its contraction dimension is the small element-local size~$n$.
Our principal target is a batch of independent small tensors, such as those arising in an element loop, rather than a single large dense matrix product.

Applying an emulated GEMM to each mode product requires encoding
the floating-point input through scaling and conversion to integer
residues, then reconstruct the result.
Even when the operator encodings are cached, these conversions
must be repeated for the changing tensor.
The chain in~\eqref{eq:intro-mode-chain} therefore requires three
tensor encodings and reconstructions, or $d$ of each for a chain
of depth~$d$.

Large matrix multiplications can amortize these conversions over
substantial matrix arithmetic, whereas small mode products provide
much less arithmetic per converted entry.
For a fixed residue count, their product work scales as $O(n^4)$,
while tensor conversion traverses $O(n^3)$ entries.
Repeated encoding and reconstruction can consequently consume a
substantial fraction of execution time.
Fusion within an emulated GEMM reduces intermediate traffic and
launch overhead~\cite{lu2026emugemm}, but composing such GEMMs
still requires a floating-point intermediate between mode products.
This motivates keeping the intermediate tensor in residue form,
with encoding only at the beginning of the chain and reconstruction
only at the end.

To avoid repeated conversion, we can keep the intermediate tensor
in residue form and reindex its residue planes for successive
mode products without reconstructing an FP64 tensor between them.
However, the represented integer can grow as the chain progresses,
even though each stored residue remains small.
Ozaki scheme~II already discusses products of more than two matrices
by allocating a range budget for the complete
product~\cite{ozaki2025scheme2}.
For a fixed residue basis containing $s$ moduli, with product
$P=\prod_{\ell=1}^{s}p_\ell$, a sufficient condition for
unambiguous signed reconstruction of an integer $v$ is
$2|v|<P$.
The available INT8 modulus pool contains 49 pairwise coprime
moduli whose product is approximately $2^{342}$.
Even if all these moduli were selected, the centered
reconstruction range would accommodate integer magnitudes
only up to approximately $2^{341}$.

A chain of tensor mode products can require a much larger range, because each mode product multiplies the current integer entries by another encoded operator and sums the products along one mode.
For an retained significand bits $b$,
also termed mantissa bits in Ozaki schemes related work,
suppose that the encoded input
and all effective integer operators have entries bounded
in magnitude by $2^b$.
Each mode product can then increase the largest magnitude
by a factor of at most $n2^b$.
Applying this estimate over $d$ mode products gives
\begin{equation}
  \|\bar X_d\|_{\max}
  \leq n^d2^{(d+1)b},
  \label{eq:intro-range-budget}
\end{equation}
where $\|\bar X_d\|_{\max}$ is the largest absolute entry after
the final product.
This conservative estimate follows from the triangle inequality.
Keeping the bound below $P/2$ is sufficient to guarantee
unambiguous reconstruction. For example, let $b=53$ keeping full FP64 significand,
at $n=8$, $d=6$ the bound reaches $2^{389}$,
which exceeds the centered range of the entire modulus pool.
The actual integer magnitudes can be smaller because this bound
uses only entrywise magnitude limits and does not account for
cancellation.
Nevertheless, the bound can no longer certify reconstruction
within the available range.
Enlarging the basis also increases both residue storage and the
number of residue GEMMs required for each mode product.

These observations lead to a central question: How can a chain of
tensor mode products reuse an integer residue representation while
controlling the magnitude of the represented integers and retaining
the tensor-product structure?
To address this question, we draw on established techniques in
residue-number arithmetic and approximate homomorphic
encryption~\cite{cheon2017ckks,cheon2018rns,halevi2018rns}
and propose a persistent execution method that combines rescaling
and base extension between mode products.
Rescaling divides the represented integer by selected moduli,
with rounding computed directly from its residues.
Base extension then computes residues of the rescaled integer
for the removed moduli, restoring the working basis for the
next product.
Together, these operations allow the intermediate tensor to remain
in residue form throughout the chain, with encoding only at entry
and reconstruction only at exit.

\begin{enumerate}
  \item A rescaling and base-extension algorithm for
  persistent Ozaki scheme~II execution of chained tensor
  mode products. The algorithm combines operator-dependent
  range planning, separable scale handling, and mixed-radix
  base extension to reuse a fixed working residue basis
  across the chain.
  \item A GPU implementation using CuTe DSL and INT8
  tensor cores for verification and evaluation.
  \item An experimental evaluation of numerical accuracy,
  execution time, and resident device memory, including
  comparisons with GEMMul8-based implementations that
  apply successive mode products or a cached dense
  operator~\cite{gemmul8}.
\end{enumerate}

The remainder of the paper is organized as follows.
Section~\ref{sec:background} introduces tensor mode products,
sum factorization, Ozaki scheme~II, and the residue
arithmetic used in this work.
Section~\ref{sec:related-work} discusses related work.
Section~\ref{sec:rescale-base} explains how rescaling and
base extension control integer magnitude and manage the
available representation range.
Section~\ref{sec:method-tensor-execution} applies these
operations to persistent Ozaki scheme~II execution of
tensor mode products and describes the GPU implementation.
Section~\ref{sec:evaluation} evaluates numerical accuracy,
execution time, and memory usage, including comparisons
with GEMMul8.
Section~\ref{sec:conclusion} concludes the paper.

\section{Background}
\label{sec:background}

\subsection{Tensor Mode Products and Sum-factorization}
\label{sec:background-tensor}

A tensor mode product applies a matrix along one index of a tensor.
For a third-order tensor $X\in\mathbb{R}^{n_1\times n_2\times n_3}$ and a matrix $A\in\mathbb{R}^{m_1\times n_1}$, the mode-1 product
$Z=X\times_1 A$ has entries
\begin{equation}
  Z_{ajk}=\sum_{i=1}^{n_1} A_{ai}X_{ijk},
  \qquad
  Z\in\mathbb{R}^{m_1\times n_2\times n_3}.
  \label{eq:bg-mode-product}
\end{equation}
Products along modes~2 and~3 are defined in the same way.
A mode-$r$ fiber is the vector obtained by fixing every tensor index except the index of mode~$r$. The same operator matrix acts on every fiber of that mode~\cite{kolda2009tensor}.

\begin{figure}[t]
  \centering
  \includegraphics[width=0.8\linewidth]{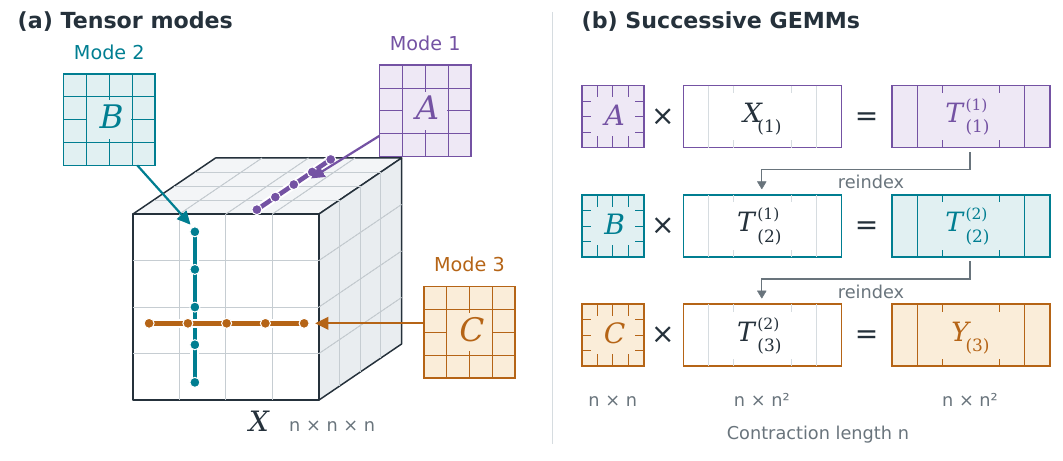}
  \caption{Three mode products and their GEMM representations.
  Panel~(a) highlights one fiber in each mode of an $n\times n\times n$
  tensor. Operators $A$, $B$, and $C$ act on all fibers of modes~1,~2,
  and~3, respectively. Panel~(b) shows the successive unfolded products
  using the intermediate tensors.
  Each GEMM applies an $n\times n$ matrix to $n^2$ fibers of length~$n$.}
  % Reindexing changes the unfolding for the next product.
  % Dimensions are schematic.
  \label{fig:mode-products}
\end{figure}

Figure~\ref{fig:mode-products} illustrates three successive mode products for square operators $A,B,C\in\mathbb{R}^{n\times n}$ and an input
$X\in\mathbb{R}^{n\times n\times n}$. We denote the intermediate tensors by
\begin{equation}
  T^{(1)}=X\times_1 A,
  \qquad T^{(2)}=T^{(1)}\times_2 B,
  \qquad Y=T^{(2)}\times_3 C.
  \label{eq:bg-three-stages}
\end{equation}
The colored fibers in panel~(a) indicate the direction of each product. 
Panel~(b) gives its matrix form. The mode-$r$ unfolding $X_{(r)}$ arranges the mode-$r$ fibers as matrix columns. With consistent column
ordering, the relation between a mode product and its unfolding is
\begin{equation}
  Z=X\times_r A
  \quad\Longleftrightarrow\quad
  Z_{(r)}=A X_{(r)}.
  \label{eq:bg-unfolding}
\end{equation}
In the square case, $X_{(r)}$ has dimensions $n\times n^2$.
Each mode product is therefore a GEMM with contraction dimension~$n$ and $n^2$ fiber columns. The next mode product unfolds the resulting intermediate tensor
along a different mode. This is a mathematical reindexing and need not
require a separate data copy in an implementation~\cite{kolda2009tensor}.

These mode products form the contraction kernels of sum factorization in high-order finite element methods. On a tensor-product hexahedral element,
a basis function is a product of three one-dimensional basis functions.
Evaluating a field at tensor-product quadrature points can therefore be performed by three successive one-dimensional interpolation operations.
For example, if $V_{ai}=\ell_i(\xi_a)$ is the one-dimensional interpolation matrix, evaluation of a coefficient tensor $U$ takes the form
\begin{equation}
  U_{\mathrm{q}}=U\times_1 V\times_2 V\times_3 V.
  \label{eq:bg-interpolation}
\end{equation}
Here $\ell_i$ is a basis function and $\xi_a$ is a quadrature point. 
The matrix $V$ may be rectangular when the number of quadrature points differs from the number of basis functions. 
Differentiation uses the corresponding derivative matrix along the required direction.
Mass and stiffness operator applications combine these contractions with quadrature weights, geometric factors, and transpose operations~\cite{swirydowicz2019tensor}.

The same three-direction operation can also be expressed using the
Kronecker product of the one-dimensional operator matrices, with their
order determined by the tensor vectorization convention~\cite{kolda2009tensor}.
This assembles the directional operators into a single matrix that acts
on the vectorized tensor, applying all three directions in one matrix
operation.
It is the assembled alternative to sum factorization, which applies
the factors successively without forming the multidimensional matrix.
For square factors, the assembled matrix has dimensions $n^3\times n^3$
and requires $O(n^6)$ storage and arithmetic for a dense application.
Sum factorization instead costs $O(n^4)$ arithmetic operations and
stores only $O(n^2)$ operator entries, together with $O(n^3)$ tensor
intermediates.

\subsection{The Ozaki Scheme II}
\label{sec:background-ozaki}

The Ozaki scheme constructs higher-precision matrix products by
splitting the operands into lower-precision components and combining
their products~\cite{ozaki2012eft}.
This approach has been implemented using both floating-point and
integer Tensor Cores~\cite{mukunoki2020tensor,ootomo2024integer}.
Ozaki scheme~II follows a different approach based on integer
residues~\cite{ozaki2025scheme2}.
After scaling and truncating the operands to integers, it represents
them by residues with respect to several pairwise coprime moduli.
The resulting residue matrices can be multiplied independently using
INT8 Tensor Cores, and their products are combined through the Chinese
remainder theorem (CRT) to reconstruct the integer result.

Consider a matrix multiplication $H=FG$, where
$F\in\mathbb{R}^{m\times k}$ and $G\in\mathbb{R}^{k\times t}$.
Ozaki scheme~II computes an approximation to $H$ in four steps.

First, the operands are scaled to retain their significant information
in an integer representation. Positive diagonal matrices $D_L$ and $D_R$, with power-of-two entries,
scale the rows of $F$ and the columns of $G$, respectively.
The scaled entries are then truncated toward zero to obtain
\begin{equation}
  \bar F=\operatorname{trunc}(D_LF),
  \qquad
  \bar G=\operatorname{trunc}(GD_R).
  \label{eq:bg-ozaki-encoding}
\end{equation}
The choice of scaling determines both the magnitudes of these integers
and the precision retained from the original operands.

Second, these integers are converted into residue matrices that fit
the low-precision multiplication units.
A basis $\mathcal P=\{p_1,\ldots,p_s\}$ of pairwise coprime moduli
is selected, and each operand is reduced modulo every modulus.
For INT8 computation, the moduli are at most 256, with a typical
basis beginning with $256,255,253,\ldots$.
The residues are stored as signed representatives centered around
zero, so they lie within the INT8 range $[-128,127]$.
This produces $s$ residue matrices for each operand, even when the
encoded integers themselves are too large to fit in INT8.

Third, each residue matrix of $\bar F$ is multiplied by the residue
matrix of $\bar G$ associated with the same modulus.
These products use INT8 Tensor Cores with INT32 accumulation.
Reducing the accumulated entries modulo that modulus gives the
residues of the integer product.
The conversion and multiplication steps can therefore be written as
\begin{equation}
  R_\ell
  =\bigl[(\bar F\bmod p_\ell)(\bar G\bmod p_\ell)\bigr]
      \bmod p_\ell,
  \qquad \ell=1,\ldots,s.
  \label{eq:bg-residue-gemm}
\end{equation}
Here, each modular reduction is applied entrywise using centered
representatives.
The products for different moduli are independent.

Finally, the output residue matrices are combined using CRT.
Let $P=\prod_{\ell=1}^{s}p_\ell$ and $P_\ell=P/p_\ell$.
Pairwise coprimality ensures that $P_\ell$ has a multiplicative inverse
modulo $p_\ell$, allowing reconstruction in the form
\begin{equation}
  \bar H=
  \left(
    \sum_{\ell=1}^{s}
    R_\ell P_\ell
    \bigl(P_\ell^{-1}\bmod p_\ell\bigr)
  \right)\bmod P.
  \label{eq:bg-crt-reconstruction}
\end{equation}
The factor $P_\ell(P_\ell^{-1}\bmod p_\ell)$ multiplying
$R_\ell$ is congruent to one modulo $p_\ell$ and zero
modulo every other modulus in the basis.
The sum therefore reproduces each output residue matrix
when reduced modulo its corresponding modulus.
The reconstructed matrix therefore matches all the output residues.
The final reduction selects the centered representative modulo $P$,
which recovers $\bar H=\bar F\bar G$ when the intended integer entries
lie within that reconstruction interval.
Removing the original scaling then gives
\begin{equation}
  \widehat H=D_L^{-1}\bar H D_R^{-1}\approx FG.
  \label{eq:bg-ozaki-unscaling}
\end{equation}

\subsection{Residue Arithmetic Used in This Work}
\label{sec:background-residues}

The residue operations used in this work build on residue number
system (RNS) arithmetic.
An RNS represents an integer by its remainders with respect to a
basis $\mathcal P$ of pairwise coprime moduli~\cite{garner1959residue}.
This representation allows addition and multiplication to be carried
out independently on smaller residues, without carry propagation
between components.
Ozaki scheme~II uses this property to compute matrix products through
low-precision integer operations.
The residue tuple determines an integer modulo the basis product
$P=\prod_{p\in\mathcal P}p$.
Recovering the intended integer therefore also requires a known
reconstruction interval.

To store these residues as signed low-precision integers, we use
centered representatives.
For an integer $v$ and modulus $p\geq2$, first compute the
nonnegative remainder $r=v-p\lfloor v/p\rfloor$.
This remainder lies in $\{0,\ldots,p-1\}$.
Subtracting $p$ from remainders in the upper part of this range gives
the centered residue
\begin{equation}
  [v]_p=
  \begin{cases}
    r, & r<\lceil p/2\rceil,\\
    r-p, & r\geq\lceil p/2\rceil.
  \end{cases}
  \label{eq:bg-centered-calculation}
\end{equation}
The resulting representative satisfies
\begin{equation}
  -\lfloor p/2\rfloor\leq[v]_p\leq\lceil p/2\rceil-1.
  \label{eq:bg-centered-residue}
\end{equation}
For example, the nonnegative remainder 5 modulo 7 becomes
$5-7=-2$ in the centered interval $[-3,3]$.
For moduli at most 256, this interval lies within the INT8 range
$[-128,127]$, allowing the residues to serve as operands in the
integer matrix products of Ozaki scheme~II~\cite{ozaki2025scheme2}.

Changing to centered representatives preserves the rules of modular
addition and multiplication.
In particular, reducing the operands before an operation gives the
same residue as reducing its result,
\begin{equation}
  [uv]_p=[[u]_p[v]_p]_p,
  \qquad
  [u+v]_p=[[u]_p+[v]_p]_p.
  \label{eq:bg-residue-operations}
\end{equation}
These identities also hold for sums of products, which explains why
tensor contractions can be computed independently for each modulus.
Since reduction acts on individual entries, reindexing a tensor
also preserves its residue representation.
The same residue planes can therefore be used for successive
tensor modes.

Although modular reduction keeps each stored residue small, it
does not reduce the magnitude of the integer represented by the
complete tuple.
Managing that magnitude requires a different operation.
We use rescaling to replace the represented integer by a rounded
quotient, followed by base extension to represent that quotient
on a larger basis.

\subsubsection{Rescaling}
\label{sec:background-rescaling}

Rescaling reduces the magnitude of the represented integer through
rounded division.
Related operations are used in approximate homomorphic encryption
and its RNS implementations~\cite{cheon2017ckks,cheon2018rns}.
Suppose that $v$ is represented on $\mathcal P$ and that a modulus
$q\in\mathcal P$ is selected for removal.
Its centered residue $r_q=[v]_q$ provides the remainder needed
to form the integer quotient
\begin{equation}
  v'=\frac{v-r_q}{q}.
  \label{eq:bg-rescale}
\end{equation}
Because $|r_q|\leq q/2$, this quotient satisfies
\[
  \left|v'-\frac{v}{q}\right|\leq\frac12.
\]
Thus, subtracting the centered remainder before division produces
a nearest-integer approximation to $v/q$.

Moreover, the residues of this quotient can be obtained directly from the
residues of $v$.
For every remaining modulus $p\ne q$, pairwise coprimality ensures
that the modular inverse $q^{-1}\bmod p$ exists.
Reducing the quotient relation modulo $p$ therefore gives
\begin{equation}
  [v']_p
  =\bigl[([v]_p-r_q)(q^{-1}\bmod p)\bigr]_p.
  \label{eq:bg-rescale-residue}
\end{equation}
This update applies only to the remaining moduli because $q$ has
no multiplicative inverse modulo itself.
The quotient is consequently represented on the reduced basis
$\mathcal P'=\mathcal P\setminus\{q\}$, whose product is $P'=P/q$.
Rescaling therefore reduces both the integer magnitude and the
available representation range.
To restore the original basis, we must compute the missing residue
$[v']_q$, which is the role of base extension.

\subsubsection{Base Extension}
\label{sec:background-base-extension}

Base extension computes residues of the same represented integer
on additional coprime moduli~\cite{halevi2018rns}.
After the rescaling step above, it restores $\mathcal P$ by
computing $[v']_q$ from the residues on $\mathcal P'$.
The integer remains $v'$, while the additional modulus enlarges
the range available for subsequent operations.

For this extension to represent the intended quotient, the remaining
residues must identify $v'$ within a known reconstruction interval.
For centered recovery, $2|v'|<P'$ is a sufficient condition.
Integers differing by a multiple of $P'$ have identical
residues on $\mathcal P'$ but can have different residues modulo $q$.
Without a known reconstruction interval, the remaining residues
alone cannot distinguish those possibilities.

Mixed-radix conversion provides one way to compute the additional
residue.
Order the remaining moduli as
$\mathcal P'=\{\widetilde p_1,\ldots,\widetilde p_{s-1}\}$.
The nonnegative CRT representative $u\in[0,P')$ of the quotient
residues has the mixed-radix expansion
\begin{equation}
  u=\sum_{i=1}^{s-1}a_i W_i,
  \qquad W_1=1,
  \qquad W_i=\prod_{j<i}\widetilde p_j,
  \qquad 0\leq a_i<\widetilde p_i.
  \label{eq:bg-mixed-radix}
\end{equation}
The digits $a_i$ can be obtained by successive modular elimination
using Garner's method~\cite{garner1959residue}.
Once these digits are known, the expansion can also be evaluated
in nested form using Horner's method~\cite{horner1819method}.
Starting from the last digit, each step multiplies the current
value by the preceding modulus and adds the corresponding digit.
Performing this evaluation modulo $q$ gives $u\bmod q$.

The nonnegative representative $u$ may differ from the intended
signed quotient.
To account for this, write $v'=u-\delta P'$, where $\delta=0$
for $u<P'/2$ and $\delta=1$ otherwise.
Reducing this expression modulo $q$ gives
\begin{equation}
  [v']_q=
  \left[
    \sum_{i=1}^{s-1}a_i(W_i\bmod q)
    -\delta(P'\bmod q)
  \right]_q.
  \label{eq:bg-base-extension}
\end{equation}
Appending this residue restores the basis $\mathcal P$.
The extended tuple still represents the rescaled integer $v'$,
so base extension restores representation capacity but does not
recover precision lost during rescaling.

\section{Related Work}
\label{sec:related-work}

\subsection{Tensor-product Kernels in Finite Element Methods}
\label{sec:related-fem}

GPU implementations of tensor-product finite element kernels have
progressed from optimizing data movement on conventional GPU cores
to exploiting Tensor Cores.
\'{S}wirydowicz et al.~\cite{swirydowicz2019tensor} study mass and
stiffness operator applications and examine how shared-memory use,
intermediate storage, and data movement affect performance.
Their work provides a foundation for organizing successive
element-local contractions within GPU kernels.

More recent work maps these contractions directly to Tensor Cores.
Cui~\cite{cui2024tensor} investigates tensor-product finite element
operations on NVIDIA A100 GPUs, emphasizing direct instruction-level
programming and shared-memory layouts that avoid bank conflicts.
The study considers both double-precision kernels and a
mixed-precision multigrid-preconditioned Poisson solver.
Tu et al.~\cite{tu2026fp64} extend FP64 Tensor Core acceleration to
high-order finite element operators in MFEM.
They combine tensor-core programming with kernel fusion and
data-layout optimizations, and evaluate the resulting kernels
within a production wave-propagation application at large GPU scale.

These studies demonstrate the importance of matching small
tensor contractions to the hardware and reusing intermediate data.
Our implementation also targets these tensor contractions,
while addressing the representation changes introduced by
Ozaki-II emulation.
We retain intermediate tensors as integer residues across mode
products and use rescaling and base extension to manage their
representation range.
The focus is therefore on reducing repeated conversion while
preserving the factored contraction structure.

\subsection{Precision Emulation on Matrix Engines}
\label{sec:related-emulation}

The Ozaki scheme builds on error-free transformations of matrix
multiplication developed by Ozaki et al.~\cite{ozaki2012eft}.
Subsequent work adapted this approach to the low-precision matrix
units available on GPUs.
Mukunoki et al.~\cite{mukunoki2020tensor} investigated DGEMM using
floating-point Tensor Cores, including accurate and reproducible
variants, while Ootomo et al.~\cite{ootomo2024integer} developed
DGEMM emulation on integer matrix multiplication units.
Together, these studies established how exact low-precision
component products can support higher-precision matrix arithmetic.

Ozaki scheme~II introduced integer modular arithmetic into this
GEMM-oriented emulation framework~\cite{ozaki2025scheme2}.
GEMMul8 implements this approach through a floating-point matrix
interface and allows preprocessing to be reused for unchanged
operands~\cite{gemmul8}.
Such caching is useful for constant finite element operators.
However, composing these GEMM calls still requires the changing
tensor to pass through the floating-point interface between mode products,
motivating our investigation of persistent residue representations.

Recent work has improved both the efficiency and the capabilities
of Ozaki-based emulation.
To reduce intermediate data movement, EmuGEMM fuses operations
within an emulated GEMM~\cite{lu2026emugemm}.
Its Scheme-I implementation combines component products and
reconstruction in a persistent kernel.
For Scheme~II, it places modular reduction in the GEMM epilogue,
avoiding a separate global-memory write of the INT32 accumulators
before producing output residues.
It also provides a fused complex kernel.
These optimizations can support efficient residue products within
our method, while our rescaling and base extension
manage the representation passed between successive mode products.

Another direction expands the precision formats available for
emulation.
Mukunoki~\cite{mukunoki2025fp8} investigated FP8 Tensor Cores for
DGEMM, and Uchino et al.~\cite{uchino2026fp8} developed an FP8
quantization approach for Ozaki-II.
Hayashi et al.~\cite{hayashi2026fp4} extended both Ozaki schemes to
FP4 Tensor Cores by representing integers with base-13 FP4 limbs
and keeping the relevant partial products exact under FP32
accumulation.
Alongside these hardware-oriented developments, Schwarz
et al.~\cite{schwarz2025guaranteed} addressed accuracy selection
through Automatic Dynamic Precision.
Their method selects the decomposition size using an exponent-span
estimate and incorporates exception handling and fallback to native
FP64, improving the reliability of emulated GEMM across different
inputs.

Ozaki-II has also been extended beyond real matrix multiplication.
Uchino et al.~\cite{uchino2025complex} developed a Karatsuba-based
formulation for complex matrix multiplication that combines real
partial products in the residue domain before CRT reconstruction.
This provides an existing example of combining several matrix
products before reconstructing one logical output.
Our setting extends the lifetime of the residue representation
across successive tensor mode products, where each intermediate
becomes an operand of the next contraction.

A related application is the three-dimensional FFT studied by
Matsuoka~\cite{matsuoka2026fft}.
The proposed design uses Bailey decompositions and combines
emulated DFT GEMMs, fused twiddle factors, and CRT reconstruction
using Tensor Cores and fixed-point accumulation.
Its analysis identifies conversion and integer-epilogue costs as
important performance constraints.
In the described algorithm, the intermediate between the two
Bailey GEMM stages is reconstructed and then converted to residues
again.
This shares our motivation to reduce the cost of precision
emulation in structured multidimensional computations.
We instead investigate carrying the residue representation across
tensor mode products through rescaling and base extension.
The FFT study presents projected performance bounds, whereas our
evaluation reports hardware measurements.

Our method builds on the scaling rules, residue products, and CRT
reconstruction of Ozaki scheme~II.
The original paper already considers products of more than two
matrices and derives a sufficient CRT range condition for the
combined integer result~\cite{ozaki2025scheme2}.
Our contribution concerns maintaining a reusable residue
representation throughout a chain of tensor mode products.
We combine established RNS rescaling and base-extension operations
with operator-dependent range planning and separable scale handling
to control intermediate integer growth while preserving the
factored tensor computation.

\subsection{Residue Arithmetic in Exact Computation and Homomorphic Encryption}
\label{sec:related-residues}

Residue representations and mixed-radix conversion have a long history.
Garner investigated residue number systems as a means of accelerating
integer arithmetic through independent addition and multiplication
on smaller residues, without carry propagation between
components~\cite{garner1959residue}.
He also described a method for converting a residue tuple into a
mixed-radix representation.
This conversion, commonly known as Garner's method, is identified
in Ozaki scheme~II as an alternative to direct CRT
reconstruction~\cite{ozaki2025scheme2}.

Modular representations have also been used in other computational
settings.
FFLAS-FFPACK implements dense linear algebra over word-size prime
fields using optimized numerical kernels~\cite{dumas2008fflas}.
Residue-based accelerators such as Mirage use these representations
to address precision limitations in photonic DNN
computation~\cite{mirage2023}.
These studies illustrate how modular arithmetic can adapt numerical
computations to the capabilities of the underlying hardware.

Managing the scale and basis of residue representations is
particularly important in homomorphic encryption.
The CKKS scheme uses rescaling to control scale growth during
approximate arithmetic~\cite{cheon2017ckks}, and its full RNS variant
develops modulus-switching procedures that operate on residue
components~\cite{cheon2018rns}.
The RNS variants developed by Bajard et al.~\cite{bajard2016rns}
and Halevi et al.~\cite{halevi2018rns} for the
Fan--Vercauteren (FV) scheme, also known as
Brakerski--Fan--Vercauteren (BFV), provide further techniques
for scaling and basis conversion.

These works provide the arithmetic foundations for our range
management.
We apply rounded modulus division and mixed-radix base extension
to the intermediate integers of a tensor mode-product chain.
Our contribution lies in organizing these operations around the
chain's magnitude bounds and scaling requirements, allowing the
residue representation to persist between contractions.

\section{Representation Range Management through Rescaling and Base Extension}
\label{sec:rescale-base}

\subsection{A Scalar Example}

\begin{figure*}[t]
  \centering
  \includegraphics[width=\linewidth]{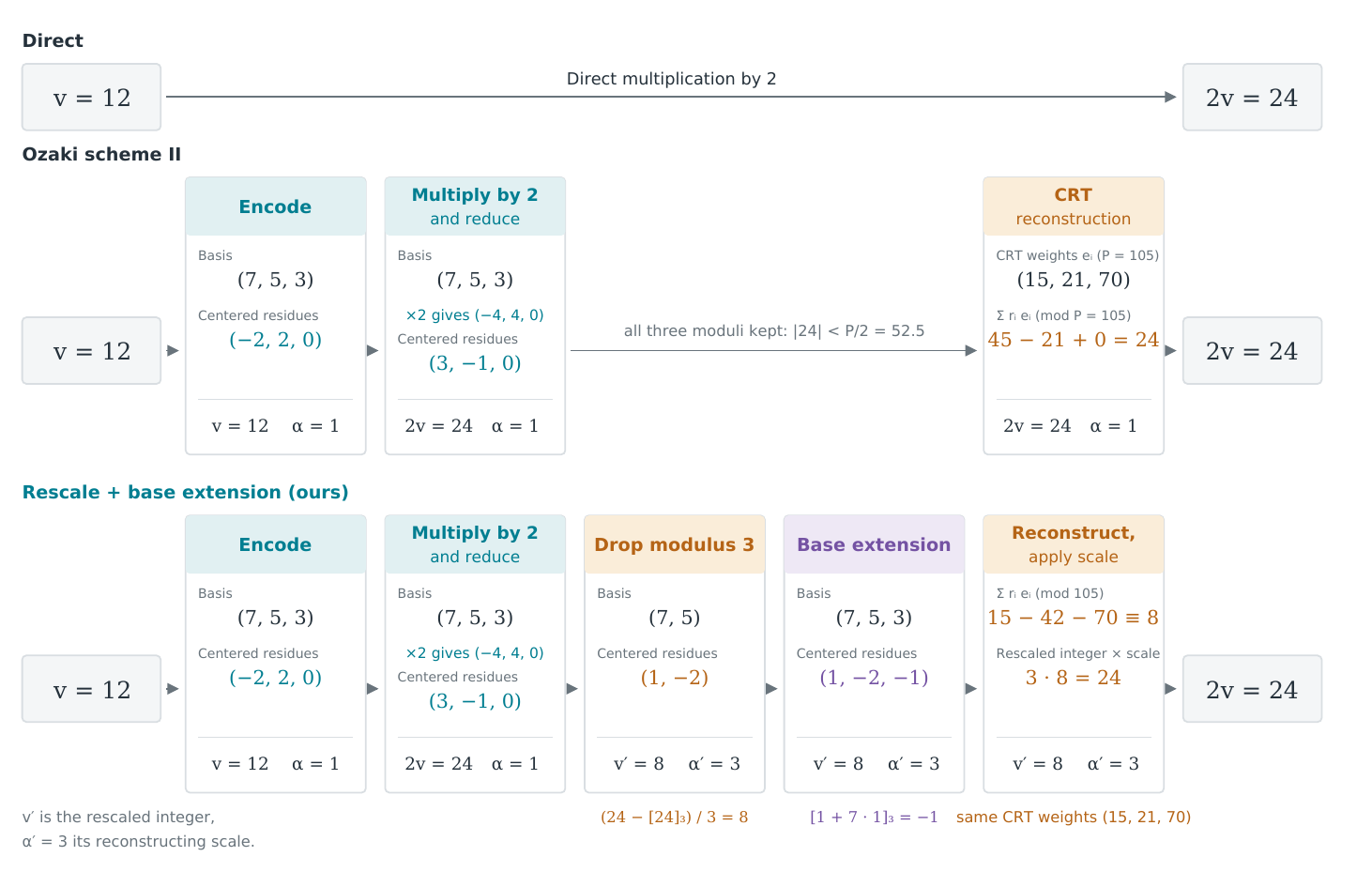}
  \caption{Three execution paths for multiplying $v=12$ by two.
  The upper path computes $2v=24$ directly.
  The middle path illustrates Ozaki scheme~II, which encodes
  the input, multiplies its residues, and reconstructs $24$
  using CRT.
  The lower path additionally applies rescaling and base extension
  before final reconstruction.
  Each residue tuple follows the displayed basis order.
  Dropping modulus $3$ changes the represented integer from $24$
  to $8$ and the reconstruction scale from $\alpha=1$ to $\alpha'=3$.
  Base extension preserves this integer and scale.
  Final reconstruction and scaling recover $24$ exactly in this
  example.}
  \label{fig:scalar-rescale-extension}
\end{figure*}

To illustrate how rescaling and base extension work together,
Figure~\ref{fig:scalar-rescale-extension} compares three paths:
direct multiplication, Ozaki scheme~II, and residue-domain
execution with rescaling and base extension.
All three paths start from $v=12$ and produce $2v=24$.

The residue paths also track a reconstruction scale $\alpha$.
Multiplying the integer recovered from the residues by this scale
gives the represented numerical value.
The initial scale normally accounts for the scaling applied during
input encoding.
Here, the input is already an integer, so we begin with $\alpha=1$.
When rescaling divides the represented integer by a modulus $q$,
the reconstruction scale is multiplied by $q$ to compensate for that
division, apart from any rounding error.

Both residue paths first encode $v=12$ on the ordered basis
$(7,5,3)$, whose product is $105$.
Its centered residues are $(-2,2,0)$.
Multiplying each residue by two gives $(-4,4,0)$, which becomes
$(3,-1,0)$ after centered reduction.
This tuple represents the doubled integer $24$, with the
reconstruction scale still equal to one.

In the Ozaki scheme~II path, all three moduli are retained,
and CRT reconstruction recovers the integer directly.
For the ordered basis $(7,5,3)$, the CRT weights are
$(15,21,70)$.
Each weight is congruent to one modulo its associated
modulus and zero modulo the other two.
The reconstructed result is therefore
\[
  [3\cdot15+(-1)\cdot21+0\cdot70]_{105}
  =[45-21]_{105}=24.
\]
Since $2|24|<105$, centered reconstruction recovers the
intended integer.
The reconstruction scale remains $\alpha=1$, so the
numerical result is also $24$.

The rescaling and base-extension path instead reduces
the represented integer before final reconstruction.
It first drops the modulus $q=3$.
Since the corresponding residue is $[24]_3=0$, the rescaled
integer is
\begin{equation}
  v'=\frac{2v-[2v]_3}{3}=\frac{24-0}{3}=8.
  \label{eq:method-scalar-drop}
\end{equation}
This expression describes the change in the represented integer.
In residue arithmetic, the remaining residues are computed directly
using~\eqref{eq:bg-rescale-residue}, without reconstructing $24$.
The result is $(1,-2)$ on the reduced basis
$\mathcal P'=\{7,5\}$, whose product is $P'=35$.
At the same time, the reconstruction scale changes from $\alpha=1$
to $\alpha'=\alpha q=3$.
The smaller integer and the larger scale therefore still represent
the numerical value $3\cdot8=24$.

The reduced basis is sufficient to identify $v'=8$ because
$2|v'|=16<P'$.
Base extension can therefore recover its missing residue modulo
$3$.
Using the centered mixed-radix expansion $8=1+7\cdot1$ gives
$[1+7\cdot1]_3=-1$.
Appending this residue produces $(1,-2,-1)$ on the original basis
$(7,5,3)$.
The represented integer remains $8$, and the reconstruction scale
remains $3$.
As shown in the figure, the same CRT weights now recover
\[
  [1\cdot15+(-2)\cdot21+(-1)\cdot70]_{105}
  =[-97]_{105}=8.
\]
Applying the reconstruction scale then gives
\[
  \alpha'v'=3\cdot8=24=2v.
\]

The recovery is exact in this example because $24$ is divisible
by the dropped modulus $3$, so rescaling introduces no rounding
error.
This is not true for every input.
Starting instead from $v=13$ gives a doubled integer of $26$,
whose centered residue modulo $3$ is $-1$.
Rescaling therefore produces $(26-(-1))/3=9$.
With the updated scale of $3$, the decoded result is $27$ rather
than $26$.
Base extension preserves the rescaled integer and cannot recover
the precision lost in this rounding step.

\subsection{Integer Growth and Reconstruction Range}

In the scalar example, the represented integers remain within
the centered reconstruction range at every step.
For a chain of mode products, this condition must be maintained
as the integer magnitudes change.
The representation range is an essential constraint in Ozaki
scheme~II.
For a residue basis with product $P$, CRT determines an integer
only modulo $P$.
Centered reconstruction selects a representative from an interval
around zero, with a magnitude limit of approximately $P/2$.
If the intended integer falls outside this interval, reconstruction
returns a different integer with the same residues.
The stored residues can therefore remain valid even when the
intended value can no longer be recovered through centered
reconstruction.

This constraint must hold throughout a persistent mode-product
chain.
The input is encoded once, and each subsequent product updates
the represented integers without reconstructing an intermediate
floating-point tensor.
Multiplication by an encoded operator and summation along a tensor
mode can increase their magnitudes, so a basis that accommodates
the first product may be insufficient for later products.
Reducing each stored residue modulo its own modulus does not
prevent this growth.

\begin{figure*}[t]
  \centering
  \includegraphics[width=0.8\linewidth]{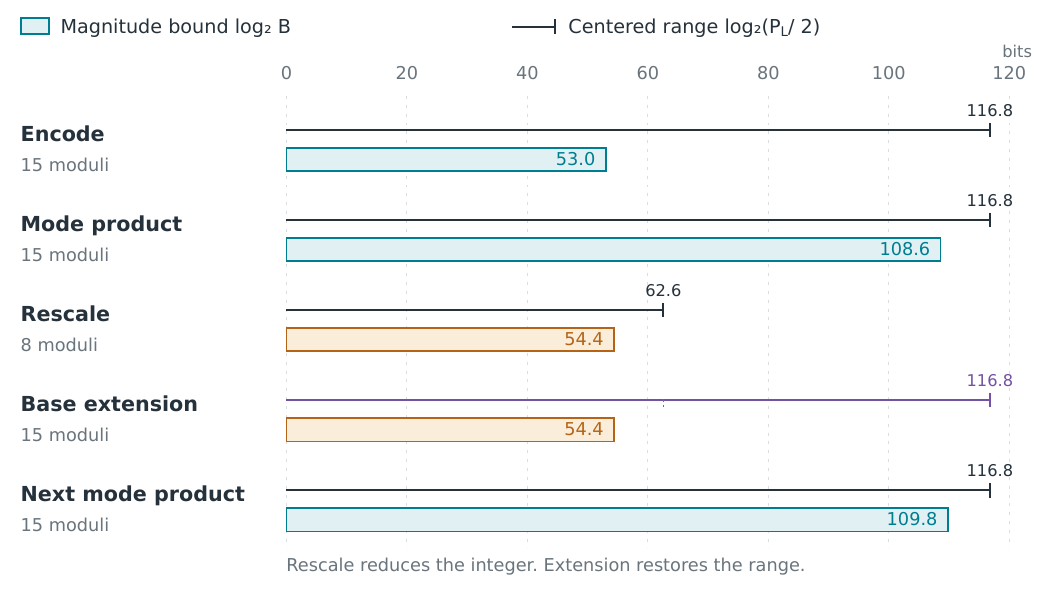}
  \caption{Changes in the integer magnitude bound and the available
  centered reconstruction range.
  Filled bars show $\log_2 B$, and capped lines show
  $\log_2(P_{\mathcal L}/2)$ for the currently available basis.
  Rescaling removes seven moduli and reduces both quantities.
  Base extension restores the original range without changing the
  rescaled integer.
  The displayed exponents are rounded illustrative values,
  not measurements.}
  \label{fig:range-rescale-extension}
\end{figure*}

Figure~\ref{fig:range-rescale-extension} illustrates how rescaling
and base extension manage the relationship between integer
magnitude and representation range.
Let $B$ be an upper bound on the absolute value of every represented
integer entry.
Let $\mathcal P$ denote the working residue basis and
$\mathcal L\subseteq\mathcal P$ the currently available basis,
whose product is $P_{\mathcal L}=\prod_{p\in\mathcal L}p$.
A sufficient condition for recovering the intended integers is
\begin{equation}
  2B<P_{\mathcal L}.
  \label{eq:method-range}
\end{equation}

The figure shows both quantities on a logarithmic scale.
The filled bars represent the magnitude bound $\log_2 B$, while
the capped lines indicate the centered range limit
$\log_2(P_{\mathcal L}/2)$.
The working basis contains 15 moduli from the INT8 Ozaki-II table,
which begins with $256,255,253,\ldots$.
Their product satisfies $\log_2 P\approx117.8$, giving an initial
centered range limit of approximately 116.8 on this scale.
The initial magnitude bound is determined separately by the
significand bits $b$, which bounds the encoded entries by $2^b$.
With $b=53$, the figure therefore starts from $\log_2 B=53$.

A mode product can increase the magnitude bound while leaving
the representation range unchanged.
Rescaling reduces this magnitude by dividing the represented
integer by a selected modulus $q$, with rounding.
It also removes $q$ from the available basis, reducing the basis
product from $P_{\mathcal L}$ to $P_{\mathcal L}/q$.
The figure illustrates seven successive modulus drops, which
reduce both the magnitude bound and the available range.

Base extension then computes residues of the rescaled integer
for the removed moduli, restoring the working basis $\mathcal P$.
The integer remains unchanged, but the restored basis provides
additional range for the next product.
Together, rescaling and base extension reduce the integer
magnitude while restoring representation capacity between
successive mode products.
Base extension does not reverse the rounding introduced by
rescaling.

\subsection{Rescaling and Drop planning}

The procedure \texttt{RnsDrop} in Algorithm~\ref{alg:rnsdrop}
applies the rescaling update from
Section~\ref{sec:background-rescaling} to every tensor entry.
For a selected modulus $q$, each remaining residue plane is
updated using the corresponding entry from the $q$ plane and
a precomputed modular inverse.
If an entry represents an integer $v$ before the drop, the update
represents $v'=(v-r_q)/q$ on $\mathcal L\setminus\{q\}$,
where $r_q=[v]_q$.
The reconstruction scale changes from $\alpha$ to $\alpha'=\alpha q$,
giving the local error
\begin{equation}
  \alpha'v'-\alpha v=-\alpha r_q,
  \qquad
  |\alpha'v'-\alpha v|\leq |\alpha|q/2.
  \label{eq:method-rescale-error}
\end{equation}
% During execution, \texttt{RnsDrop} is applied after a mode product
% whenever the drop schedule requires rescaling.

\RnsDropAlgorithm

To determine the modulus-dropping schedule for the rescaling phase before execution, 
we first estimate how much each mode product can increase the integer magnitude.
Let $\bar A_i$ denote the effective integer operator for the $i$-th mode product,
including the scaling incorporated into that operator.
Each output entry of the mode product is an inner product
between one row of $\bar A_i$ and a tensor fiber.
Let $j$ index the output entries along that fiber and $k$
index the entries being contracted.
The maximum absolute row sum of the operator is
\begin{equation}
  \rho_i=\max_{1\leq j\leq n}
         \sum_{k=1}^{n}|(\bar A_i)_{jk}|.
  \label{eq:method-row-bound}
\end{equation}
For an integer fiber $x$ whose entries satisfy $|x_k|\leq B$,
the triangle inequality gives
\begin{equation}
  |(\bar A_i x)_j|
  \leq \sum_{k=1}^{n}|(\bar A_i)_{jk}|\,|x_k|
  \leq B\sum_{k=1}^{n}|(\bar A_i)_{jk}|
  \leq B\rho_i.
  \label{eq:method-row-bound-application}
\end{equation}
Since the same operator acts on every fiber along the contracted
mode, $B\rho_i$ bounds every entry of the output tensor.

The planner can therefore update the magnitude bound by
$B\gets B\rho_i$ at each mode product.
This is the standard induced infinity-norm bound, with
$\rho_i=\|\bar A_i\|_\infty$.

Using these operator bounds, \texttt{DropPlan} in
Algorithm~\ref{alg:dropplan} determines how many moduli to drop
after each mode product.
It starts from $B=2^b$, where $b$ is the significand bits,
so that every encoded input integer has magnitude at most $2^b$,
and updates $B\gets B\rho_i$ for each mode product.
The updated bound must satisfy $2B<P_{\mathcal L}$ on the
currently available basis.

\DropPlanAlgorithm

After each nonfinal product, the planner considers dropping
the smallest available modulus $q$.
It schedules a drop when $\lfloor B/q\rfloor\geq2^b$ and
uses the conservative bound
\begin{equation}
  B'=\lfloor B/q\rfloor+1
  \label{eq:method-drop-bound}
\end{equation}
for the rescaled integer.
The additional unit accounts conservatively for rounding
during division.
Before accepting the drop, the planner checks
$2B'<P_{\mathcal L}/q$ so that the quotient remains recoverable
on the reduced basis.
It then updates the bound and the available basis and considers
the next smallest modulus.
This process continues while the threshold is satisfied and
more than one modulus remains.
A failed range check causes the candidate schedule to be rejected.

This selection is greedy.
At each step, the planner chooses the smallest available modulus
rather than comparing alternative drop sequences.
For our proposed method, the restoration flag is set to $\tau=1$.
The planner therefore assumes that base extension restores the
working basis after the scheduled drops and uses that basis
when checking the next product.
The flag allows the same planner to support the rescale-only
variant in Appendix~\ref{app:OzII-Rescale}, which uses $\tau=0$.
When applying only the proposed method, $\tau$ remains fixed at one.
It returns a drop count $g_i$ for each mode product.
During execution, the same sequence is reproduced by calling
\texttt{RnsDrop} $g_i$ times, each time removing the smallest
available modulus, before applying base extension.
The final product is decoded directly, so $g_d=0$.

The remaining decision is the significand bits $b$.
The procedure \texttt{Planner} in
Algorithm~\ref{alg:basisplan} selects it on the supplied working
basis.
Because changing $b$ changes the encoded integer operators,
we write $\rho_i(b)$ for the corresponding maximum absolute
row sum for the $i$-th mode product.
Starting from $b=53$, which is the FP64 significand, the procedure prepares the effective
integer operators, computes their bounds, and calls
\texttt{DropPlan} with restoration enabled.
If the schedule fails a range check, it decreases $b$ by one
and repeats these steps.
It returns the first successful significand bits and its
drop counts.
If no positive $b$ passes the checks, it reports failure.
The working basis remains fixed throughout this search.

Planning depends on the operators, their contraction order,
and the working basis.
It does not depend on the individual input tensor because
the encoded input entries share the bound $2^b$.
The plan and prepared operator residues can therefore be
reused across tensors with the same operator chain.
The selected $b$ specifies the encoding precision accepted
by this planner, rather than guaranteeing $b$ accurate bits
in the final result.
In particular, rescaling can lose accuracy when individual
entries are much smaller than the propagated bound, including
when cancellation occurs.

\PlanRescaleBEAlgorithm

\subsection{Base Extension Using Shared Mixed-Radix Digits}

After the scheduled drops, base extension restores the removed
moduli while preserving the rescaled integer.
Let $\mathcal{P}'=\{q_1,\ldots,q_m\}$ be the reduced basis,
processed in the order $q_1,\ldots,q_m$, with product
$P'=\prod_{i=1}^{m}q_i$.
For each rescaled integer entry $v'$, we require $2|v'|<P'$
so that its remaining residues uniquely determine its value.
The currently available basis is $\mathcal{L}=\mathcal{P}'$,
and the moduli to be restored form the target set
$\mathcal{D}=\mathcal{P}\setminus\mathcal{P}'$.

To compute the missing residues, we first convert the remaining
residues into mixed-radix digits.
Algorithm~\ref{alg:mrc} provides the procedures
\texttt{MrcDigits} and \texttt{BaseExtend} for these two operations.
For each tensor entry, \texttt{BaseExtend} calls
\texttt{MrcDigits} once and reuses the resulting digits for
every modulus in $\mathcal{D}$.
The conversion follows the successive modular elimination
underlying Garner's method~\cite{garner1959residue}.
Unlike the nonnegative digits introduced in
Section~\ref{sec:background-residues}, we use centered digits.
If the basis contains an even modulus, we place it first,
followed by the odd moduli.
The descending basis used here satisfies this ordering when
it contains modulus $256$.

In this representation, each digit $d_i$ has a positional
weight $W_i$, analogous to the powers of ten in decimal
notation.
For mixed-radix digits, the weight is the product of all
preceding moduli,
\begin{equation}
  W_1=1,
  \qquad
  W_i=\prod_{j<i}q_j.
\end{equation}
The rescaled integer is then expressed as
\begin{equation}
  v'=\sum_{i=1}^{m}d_iW_i,
  \qquad
  -\lfloor q_i/2\rfloor\leq d_i\leq\lceil q_i/2\rceil-1.
  \label{eq:method-centered-mrc}
\end{equation}
Successive modular elimination produces digits whose weighted
sum is congruent to $v'$ modulo every modulus in $\mathcal{P}'$.
With the stated ordering, the possible weighted sums cover
exactly the centered interval
$[-\lfloor P'/2\rfloor,\lceil P'/2\rceil-1]$.
Since the intended integer also lies in this interval,
uniqueness guarantees that the expansion equals $v'$.
The modulus ordering is therefore required for this centered
representation.

Once the digits are available, each missing residue can be
computed by reducing the same expansion modulo its target
modulus.
For every removed modulus $p\in\mathcal{D}$, we precompute
the reduced positional weights $\omega_{i,p}=[W_i]_p$
and evaluate
\begin{equation}
  R_p=\left[\sum_{i=1}^{m}d_i\omega_{i,p}\right]_p.
  \label{eq:method-extension}
\end{equation}
Thus, the digit conversion is performed only once per tensor
entry, rather than repeated for each restored modulus.
It requires $m(m-1)/2$ modular elimination updates,
followed by $m$ weighted terms for each target modulus.
Only the reduced weights are needed during extension,
so the computation does not materialize the full integer $v'$
or reconstruct an intermediate FP64 tensor.

Appending the new residues restores
$\mathcal{L}=\mathcal{P}$ and its product $P$.
The existing residues, the reconstruction scale $\alpha'$, and
the rescaled magnitude bound $B'$ remain unchanged.
The restored residue planes are then available for the next
mode product, subject to the range condition checked by
the planner.

\BaseExtensionAlgorithm

% \section{Persistent execution of tensor mode products}
\section{Applying Ozaki Scheme~II to Tensor Mode Products}
\label{sec:method-tensor-execution}

This section combines the residue operations with Ozaki scheme~II
to execute chained tensor mode products.
It first describes the composition of separate Ozaki scheme~II
mode products, then explains how rescaling and base extension
allow intermediate tensors to remain in residue form across
the chain.
Finally, it describes the GPU implementation of these algorithms.

\subsection{Composed Ozaki-II Execution}
\label{sec:method-composed}

\begin{figure*}[t]
  \centering
  \includegraphics[width=0.8\linewidth]{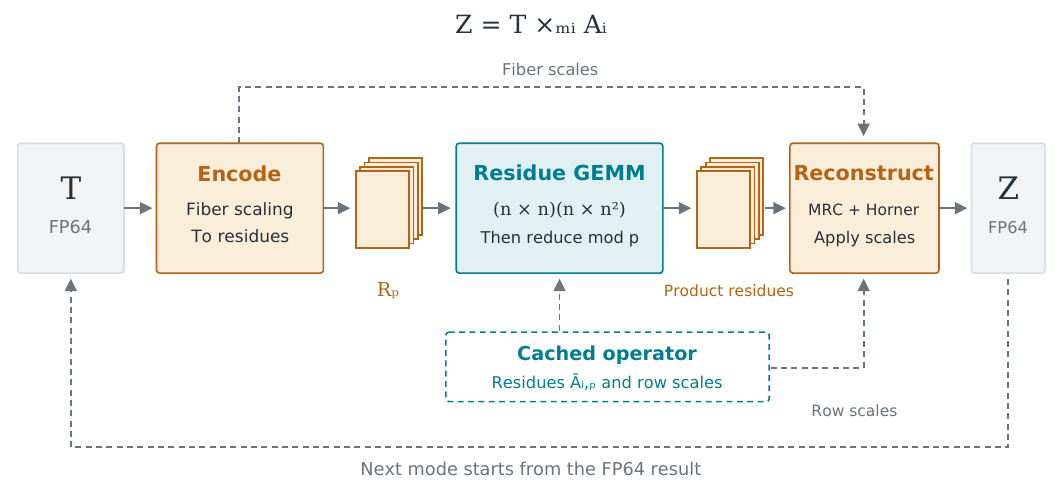}
  \caption{One tensor mode product in OzII-Composed.
  The input tensor is encoded, multiplied independently on
  each residue plane, and decoded to FP64.
  Operator residues can be cached when the operator and
  encoding settings remain unchanged.
  Each modulus requires an $(n\times n)(n\times n^2)$ GEMM.
  The next mode product repeats the conversion for the
  FP64 intermediate.}
  \label{fig:composed-ozaki2}
\end{figure*}

Figure~\ref{fig:composed-ozaki2} illustrates one mode product
in OzII-Composed.
For the $i$-th mode product, unfolding along mode $m_i$
arranges the $n^2$ fibers as columns of an $n\times n^2$
matrix.
Each fiber receives a power-of-two encoding scale, and each
operator row receives its own scale.
After scaling and rounding to integers, both integer operands are
converted to centered residue planes.
For each modulus, the operator residue matrix multiplies
the corresponding tensor unfolding, and the accumulated
result is reduced to centered residues.
Reconstruction and inverse scaling produce the FP64 tensor
used by the next mode product.
This follows the GEMM-based organization of
Ozaki scheme~II~\cite{ozaki2025scheme2}.

Algorithm~\ref{alg:encode} defines the conversion procedures.
For a group of entries $G$ that share an encoding scale,
the exponent $e_G$ is chosen so that the largest scaled
magnitude is less than $2^b$ before rounding.
An all-zero group is encoded as zero with exponent zero.
Unlike the truncation used in the original Ozaki scheme~II
formulation, our implementation rounds scaled operands to the
nearest integer, with ties to even.
The resulting integer magnitudes are bounded by $2^b$.
For the tensor input, an encoding group is one fiber in
OzII-Composed and the complete tensor in the persistent method.
The procedure \texttt{ToRns} then converts an integer tensor
$\bar T$ into one centered residue plane
$R_p=[\bar T]_p$ for each modulus $p$ in the supplied basis
$\mathcal L$, with reduction applied entrywise.

\TensorEncodingAlgorithm

For the $i$-th mode product, let $e_{i,c}$ be the encoding
exponent of fiber column $c$ and $f_{i,a}$ the encoding
exponent of operator row $a$.
If $T$ denotes the input unfolding and $k$ indexes its
contracted dimension, the integer operands are
\begin{equation}
  (\bar T)_{kc}
  =\operatorname{round}(2^{e_{i,c}}T_{kc}),
  \qquad
  (\bar A_i)_{ak}
  =\operatorname{round}(2^{f_{i,a}}(A_i)_{ak}).
  \label{eq:method-composed-encoding}
\end{equation}
Every term contributing to output entry $(a,c)$ therefore
has the same combined scale.
After reconstruction, that entry is multiplied by
$\sigma_{i,ac}=2^{-f_{i,a}-e_{i,c}}$ to undo the encoding
scales.
Together, these factors form an entrywise reconstruction scale
field determined by the operator rows and tensor fibers.

Algorithm~\ref{alg:OzIIcomposed} repeats these steps over
the chain.
The supplied basis $\mathcal P$ contains the first $s$
entries in GEMMul8's table of pairwise coprime moduli,
starting with $256$.
With $P=\prod_{p\in\mathcal P}p$, the sufficient condition
$n2^{2b}<P/2$ ensures that each length-$n$ integer dot
product remains within the centered reconstruction range.
The algorithm selects the largest positive integer
$b\leq53$ satisfying this condition and uses it to encode
both operands.
The basis remains unchanged throughout execution.

For fixed operators and encoding settings, the operator
residues are prepared once and reused.
The intermediate tensor, however, is encoded and decoded
for every mode product.
A three-direction pass therefore requires three tensor
encodings and three reconstructions.

\TensorComposedAlgorithm

To perform the reconstruction, \texttt{ReconstructFP64} uses centered
mixed-radix conversion followed by FP64 Horner evaluation
and scaling.
This reconstructs the integer represented by the residues
without evaluating the direct CRT formula used in
Ozaki scheme~II.
Mixed-radix conversion is also used for base extension
in our persistent method, so this choice allows both
operations to share the same conversion procedure.
The composed variant uses the same reconstruction procedure
for consistency.
The digit conversion is exact, while FP64 Horner evaluation
and subsequent scaling can introduce rounding errors.
The conversion in Algorithm~\ref{alg:mrc} operates on the
ordered basis supplied to the procedure, which is the
full working basis for OzII-Composed. In the pseudocode, $\operatorname{fl}_{64}$ denotes
floating-point evaluation in FP64, and $\odot$ denotes
entrywise multiplication by the supplied reconstruction scale
field $\sigma$.
The range condition guarantees a unique integer for
reconstruction, although its FP64 evaluation and subsequent
scaling may introduce rounding.
The algorithms assume finite operands and representable
scales and intermediate results during floating-point
evaluation.

\subsection{Persistent Ozaki-II Execution with Rescaling and Base Extension}
\label{sec:method-persistent-chain}

Algorithm~\ref{alg:OzIIrescalebe} combines tensor mode products
with the \texttt{PlanRescaleBE}, \texttt{RnsDrop}, and
\texttt{BaseExtend} procedures.
Given the operators, their contraction order, and the working
basis $\mathcal P$, the planner selects the significand bits
$b$ and the number of drops $g_i$ after each mode product.
The effective integer operators $\bar A_i$ are then prepared
using the same scaling and rounding rules used to compute
the bounds $\rho_i(b)$.
Operators already prepared for the accepted $b$ can be reused.

The input tensor is encoded once using a single exponent $e$.
Its initial reconstruction scale is therefore the scalar
$\alpha=2^{-e}$.
The operator row scales are tracked separately and combined
into a final reconstruction scale field, denoted by $\Gamma$.
Unlike $\alpha$, this field can vary between tensor entries.
Their treatment is described below.

\TensorPersistentAlgorithm

For the $i$-th mode product, the encoded operator multiplies
each tensor residue plane along the contracted mode.
The products use centered INT8 operands with INT32 accumulation,
followed by centered modular reduction.
The accumulations must remain within the INT32 range,
in addition to satisfying the separate CRT range conditions. When $g_i>0$, the algorithm starts from the working basis
$\mathcal L=\mathcal P$ and calls \texttt{RnsDrop} $g_i$ times.
Each call removes the smallest available modulus, following
the planner's order, and updates the scalar reconstruction scale
to account for the division.
The returned scale $\alpha'$ becomes $\alpha$ for the next
operation. After the scheduled drops, the remaining basis is
$\mathcal P'=\mathcal L$.
A single call to \texttt{BaseExtend} restores the removed
moduli in $\mathcal D=\mathcal P\setminus\mathcal P'$,
reusing the mixed-radix digits across all targets.
The next mode product therefore uses the full working
basis $\mathcal P$ again.
Since $g_d=0$, the final mode product proceeds directly
to reconstruction.

\begin{figure*}[t]
  \centering
  \includegraphics[width=0.7\linewidth]{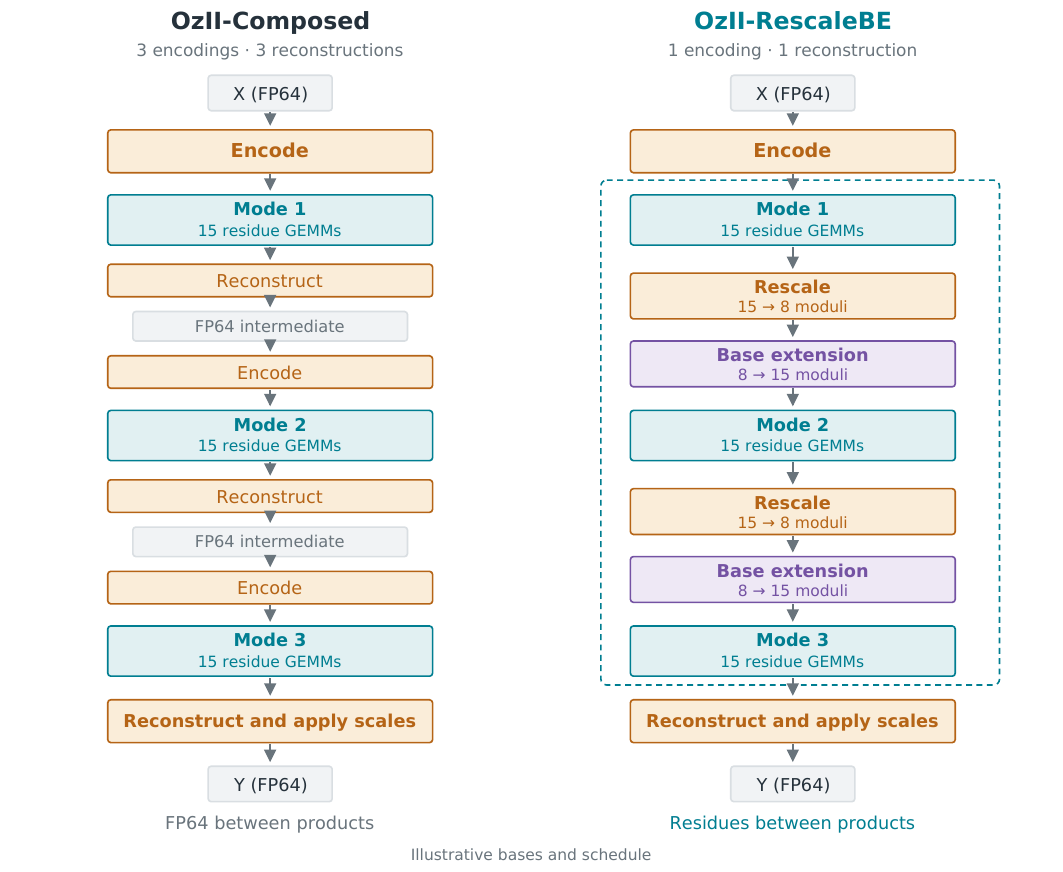}
  \caption{Three successive tensor mode products.
  OzII-Composed reconstructs an FP64 intermediate and encodes
  it again between products.
  OzII-RescaleBE retains the intermediate residues and applies
  rescaling and base extension between products.}
  % In this example, both working bases contain $15$ moduli,
  % and the persistent schedule drops and restores seven
  % moduli after each nonfinal product.
  % Both paths therefore use $45$ residue GEMMs.
  % These residue counts and drop counts illustrate one
  % configuration rather than requirements of the algorithms.}
  \label{fig:chains}
\end{figure*}

Figure~\ref{fig:chains} compares this execution with
OzII-Composed.
Both paths apply the same sequence of tensor mode products,
but handle the intermediate tensor differently.
OzII-Composed reconstructs an FP64 tensor and encodes it
again before the next product.
OzII-RescaleBE instead reduces the integer magnitude through
rescaling and restores the representation range through
base extension.
The intermediate tensor remains in residue form, accompanied
by its reconstruction scales, until the final reconstruction.

For the final reconstruction to recover the appropriate numerical
scale, the operator row scales must also be tracked throughout
the chain.
In a three-direction pass, each operator acts along a different
mode.
Its row scales depend only on the corresponding output index,
so they can be applied after all three products.
Using the row encoding exponents $f_{i,a}$ introduced for
OzII-Composed, the final operator reconstruction scale field is
\begin{equation}
  \Gamma_{a_1a_2a_3}
  =2^{-f_{1,a_1}}2^{-f_{2,a_2}}2^{-f_{3,a_3}}.
  \label{eq:method-final-operator-scale}
\end{equation}
The complete reconstruction scale is $\alpha\Gamma$.
Here, $\alpha$ accounts for the initial tensor encoding and
all modulus divisions, while $\Gamma$ accounts for the
operator row scales.

A longer chain requires an additional step when a mode is
contracted again.
The pending row scales from the previous product on that
mode now belong to the entries being summed.
They must therefore be incorporated into the next operator
before its integer encoding.
This is done by multiplying each column of that operator
by the pending reconstruction scale for the corresponding
contracted entry.

Encoding this effective operator produces new row scales,
which replace the pending scales on that mode.
The pending scales on the other modes remain unchanged.
At the end of the chain, $\Gamma$ is formed from the remaining
scale on each mode.
An earlier scale that has already been incorporated into
a later operator is therefore not applied again during
final reconstruction.

The planner computes $\rho_i(b)$ from these effective integer
operators so that its bounds describe the products actually
executed.
For each candidate $b$, operator preparation starts with
all pending scales equal to one and processes the complete
operator sequence again.
Once $b$ is selected, the resulting operator residues and
final scale field can be cached and reused for tensors
with the same operator chain.

\subsection{Implementation}
\label{sec:method-implementation}

The implementations serve as a proof of concept and use
Python, selected PyTorch utilities, and CuTe DSL version~4.5.2
for ease of development.
Python handles operator preparation, range-bound calculations,
drop planning, and the construction of modular inverse and
base-extension tables.
PyTorch provides timing and memory-measurement utilities.
Its \texttt{einsum} operation is also used to implement the
FP64 tensor mode-product baseline, using cuBLAS internally
for the tested contractions.
The GPU kernels are written in CuTe DSL and use the selected
working basis, number of significand bits, and drop schedule.
Plans and prepared operator data are reused across tensors
with the same operators and configuration.

The target GPU is the NVIDIA RTX PRO 6000 Blackwell, whose Tensor Cores support INT8
matrix multiply-accumulate (MMA) instructions.
However, CuTe DSL~4.5.2 does not expose a warp-level
INT8 MMA operation for the \texttt{sm\_120}
architecture~\cite{nvidiaCute452Warp}.
A user-defined operation using the DSL's
\texttt{@dsl\_user\_op} decorator therefore issues the
required MMA instruction through LLVM inline
assembly~\cite{nvidiaPtxMma}.

The implementations use custom residue GEMM kernels
to handle the different unfoldings required by the three
tensor mode products.
In particular, the middle-mode unfolding is not directly
available as a single contiguous matrix in the tensor's
storage layout.
The custom kernels implement the indexing for each
contracted mode and combine INT8 matrix multiplication
with centered modular reduction, reducing intermediate
data traffic.

Alongside OzII-Composed and OzII-RescaleBE, the implementation
includes OzII-NoRescale and OzII-Rescale as comparison variants.
Their algorithms and the rescale-only planner are provided
in Appendix~\ref{app:comparison-methods}.

\section{Evaluation}
\label{sec:evaluation}

\subsection{Experimental Environment}
\label{sec:experimental-environment}

Table~\ref{tab:experimental-environment} summarizes the hardware
and software environment.
The theoretical peak performance values are calculated from
NVIDIA's architecture specifications using the specified
2.28\,GHz boost clock and counting each multiply--add as
two operations~\cite{nvidiaRtxProBlackwellArchitecture}.
Tensor Core rates refer to dense operations without
structured sparsity.
The FP64 rate is calculated assuming two FP64 units per SM.

\ExperimentalEnvironmentTable

The experiments are characterized by the tensor mode size $n$,
chain depth $d$, modulus count $s$, and number of retained
significand bits $b$.
Each input tensor has dimensions $n\times n\times n$,
and each mode product applies an $n\times n$ operator along
one direction.
The chain depth $d$ counts the successive mode products,
so a pass over all three directions has $d=3$.
Longer chains revisit these directions while the tensor
remains three-dimensional.

Two baselines implement the tensor contraction using
GEMMul8~\cite{gemmul8}.
GEMMul8-Composed applies the $d$ mode products through
separate GEMMul8 calls, following the same composition
as OzII-Composed.
GEMMul8-KRON uses the Kronecker-product representation
introduced in Section~\ref{sec:background-tensor}.
For a single tensor, it multiplies a dense
$n^3\times n^3$ operator by the vectorized input of
length $n^3$.
For a batch sharing the same operator, the vectorized
inputs are collected as columns and processed in one
matrix multiplication.
This baseline produces a larger dense GEMM even at
small tensor mode sizes, providing a workload better
suited to GEMMul8's optimized matrix multiplication.
The dense operator is prepared once and reused across
evaluations.

All residue methods start from the same basis containing
$s=15$ moduli, with one residue plane per modulus.
In OzII-RescaleBE, rescaling temporarily reduces the available
basis to $\mathcal P'$, and base extension restores all $s$
moduli before the next mode product.
OzII-Rescale does not restore removed moduli.
Its final modulus counts are therefore reported alongside
the initial count.

For the OzII implementations, $b$ is selected according
to each method's range conditions on the supplied basis.
It sets the bound $2^b$ on the encoded input integer
magnitudes and, where rescaling is used, the nominal
threshold in the drop planner.
For GEMMul8, the reported $b$ is the nominal bit count
determined by the range rule, while the actual operand
scales are selected internally.
The number of significand bits $b$ is distinct from
the modulus count $s$ and does not by itself determine
the final numerical accuracy.

\subsection{Accuracy Evaluation}
\label{sec:accuracy}

\subsubsection{Datasets}
Three datasets with different input distributions are used
for the accuracy evaluation.

\textbf{Uniform.}
Each tensor entry is sampled independently and uniformly between
$-1$ and $1$.
This dataset provides an unstructured baseline without an imposed
wide exponent distribution.

\textbf{Smooth.}
Each tensor samples the field
\begin{equation}
  X(x,y,z)=a\sin(3x+\phi_1)\cos(5y+\phi_2)\exp(cz)
\end{equation}
on an $n\times n\times n$ unit grid.
For each tensor, the amplitude is sampled uniformly between 0.5 and 2,
the phases uniformly between 0 and $2\pi$, and the growth rate
uniformly between 0.5 and 1.5.
This construction provides a simple model of smooth nodal fields in high-order finite element calculations.
The analytic field is separable and has tensor rank one.
In exact arithmetic, mode products preserve this structure, and each
output entry factors into three one-dimensional contributions.
Cancellation in any contribution can produce a small output entry
and amplify its relative error.
Rounding the sampled field to FP64 can perturb exact separability.

\textbf{Wide160.}
For each entry, we sample a mantissa uniformly between 1 and 2 and
an integer exponent uniformly from $-80$ to $80$, inclusive.
The mantissa is multiplied by two raised to the sampled exponent.
The result is then assigned a positive or negative sign at random.
The name refers to the 160-bit span between the smallest and largest
exponents.
This dataset tests the effect of rounding during encoding when
entries sharing a scale have widely different magnitudes.
With a fixed $b$-bit encoding, sufficiently small entries relative to
the group's largest entry can be rounded to zero.
This applies to the per-fiber scaling in the composed methods
and the single scale per tensor in the persistent methods.
A small normwise error can therefore coexist with larger relative
errors in small output entries.

For each dataset and chain depth $d\in\{3,4,5,6,7,8\}$,
32 tensors with $n=8$ are evaluated.
Each of the $d$ operator matrices has dimensions $n\times n$,
with entries sampled independently and uniformly between
$-1$ and $1$.
The contracted modes cycle through $1,2,3,1,2,3,\ldots$.
Thus, a depth-eight chain applies three mode products
along each of modes 1 and 2 and two along mode 3.
All three datasets use the same operator distribution
and mode sequence, differing only in how the input
tensor $X$ is constructed.

\subsubsection{Error Metrics}

For each dataset and depth, the index $i$ below ranges over all
$32n^3=16384$ output entries in the test batch.

The reference $Y^{\mathrm{ref}}$ is computed by evaluating the chain using
exact rational arithmetic on the FP64 inputs and operators
with Python's \texttt{Fraction} class.
It serves as the ground truth for the evaluation.
Therefore, the FP64 results computed using PyTorch's
\texttt{torch.einsum} are also evaluated against this reference.

The normwise relative error is
\begin{equation}
  \varepsilon_{\mathrm{norm}}
  =\frac{\max_i |Y_i-Y_i^{\mathrm{ref}}|}
         {\max_i |Y_i^{\mathrm{ref}}|}.
  \label{eq:accuracy-normwise}
\end{equation}
We also compute an error for each output entry,
\begin{equation}
  \varepsilon_i=
  \begin{cases}
    |Y_i-Y_i^{\mathrm{ref}}|/|Y_i^{\mathrm{ref}}|,
      & Y_i^{\mathrm{ref}}\ne0,\\
    |Y_i-Y_i^{\mathrm{ref}}|,
      & Y_i^{\mathrm{ref}}=0,
  \end{cases}
\end{equation}
and report its maximum and median,
\begin{equation}
  \varepsilon_{\mathrm{max}}=\max_i\varepsilon_i,
  \qquad
  \varepsilon_{\mathrm{med}}=\operatorname{median}_i\varepsilon_i.
\end{equation}
% The reference $Y^{\mathrm{ref}}$ is computed by the Python's \texttt{Fraction} library, which is considered 
% can provide the ground truth values. Therefore, the FP64 reuslts calculated by the Pytorch's \texttt{torch.einsum}
% also become the evaluation target.

\subsubsection{Modulus and Encoding Configurations}

All residue methods start from the first $s=15$ moduli in
GEMMul8's ordered modulus table.
Table~\ref{tab:moduli} reports the significand bits selected
for this fixed basis at $n=8$.
The selected counts are the same for Uniform, Smooth,
and Wide160.

Ozaki scheme~II discusses successive matrix multiplications
and derives a range condition for reconstructing
their combined integer result~\cite{ozaki2025scheme2}.
As more encoded operands are multiplied, the integer result
can require a larger representation range.
On a fixed residue basis, satisfying this condition can
therefore require fewer encoding bits per operand.
The same constraint applies to a chain of tensor mode
products whose intermediate integers remain in residue form.

OzII-Composed and GEMMul8-Composed reconstruct the result
after every mode product, so their basis only needs to
accommodate one product at a time.
OzII-Composed uses $b=53$ throughout these tests.
The nominal significand bits is also $53$ for
GEMMul8-Composed and GEMMul8-KRON.
For KRON, the selected basis accommodates a single dense
product with inner dimension $n^3=512$.
GEMMul8 selects the actual operand scales internally.
GEMMul8-Composed uses its standard configuration in this
accuracy comparison.

\ModuliPerStageTable

OzII-NoRescale instead retains the integer representation
across the complete chain without rescaling.
To accommodate its growing magnitude within the same
15-modulus basis, the selected significand bits decreases
from $27$ at $d=3$ to $10$ at $d=8$.
OzII-Rescale reduces the integer magnitude between products,
but each drop also reduces the available representation
range because the removed moduli are not restored.
Its planner selects the same significand bits as
OzII-NoRescale at the tested depths.
Its final modulus counts are listed below the table.

OzII-RescaleBE reduces the integer magnitude and then restores
the working basis between products.
Its planner accepts $b=53$ at every tested depth.
For $d=3$ through $7$, seven moduli are dropped and restored
after each nonfinal mode product.
At $d=8$, seven moduli are dropped after each of the first
six products, and eight after the seventh.
Base extension restores all 15 moduli before the next
product.
These range checks permit the selected significand bits,
but do not eliminate the rounding introduced by rescaling.

\subsubsection{Accuracy Results}

\begin{figure*}[t]
  \centering
  \includegraphics[width=0.8\linewidth]{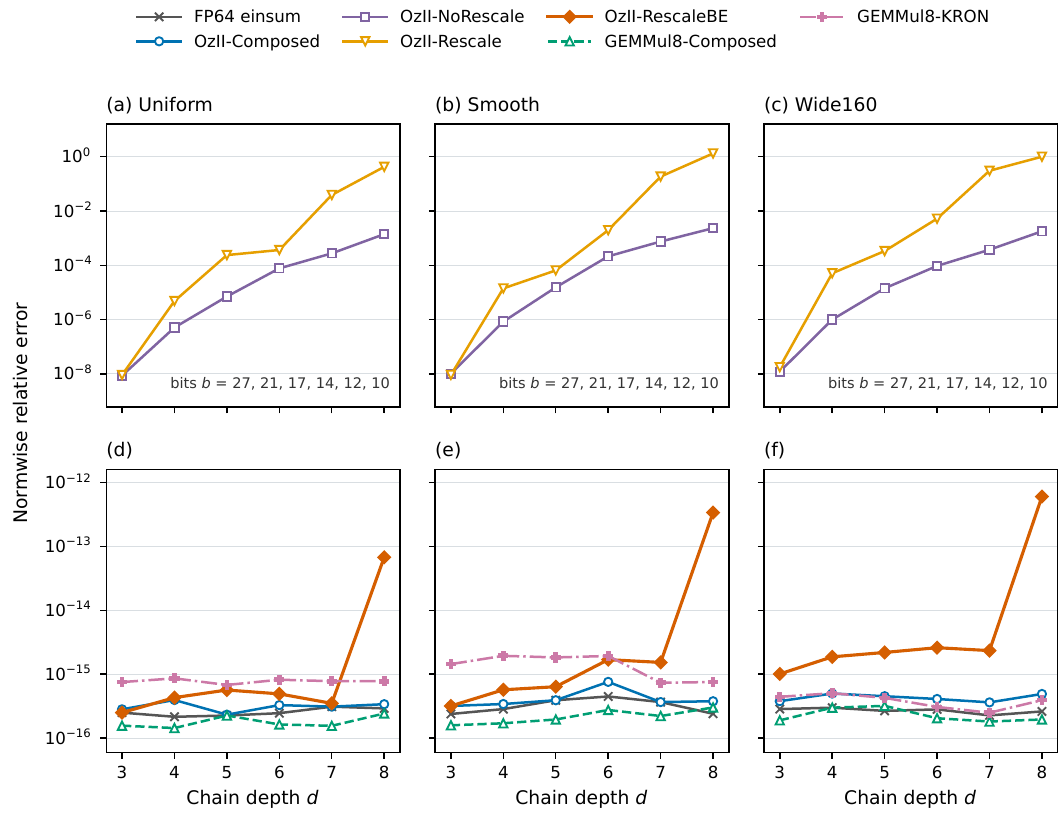}
  \caption{Normwise relative error against the reference at $n=8$.
  Columns correspond to Uniform, Smooth, and Wide160.
  Panels (a)--(c) show OzII-NoRescale and OzII-Rescale,
  with significand bits for $d=3,\ldots,8$.
  Panels (d)--(f) show the FP64 einsum and the emulation-based methods.}
  \label{fig:accuracy-depth}
\end{figure*}

Figure~\ref{fig:accuracy-depth} compares the normwise errors
across the three datasets.
The upper row shows OzII-NoRescale and OzII-Rescale, whose
significand bits decrease with depth.
The lower row shows the other residue methods together
with the FP64 einsum.
Different logarithmic vertical ranges are used for the
two rows to show their respective error levels.

OzII-NoRescale and OzII-Rescale already have normwise errors
of approximately $10^{-8}$ at $d=3$, where their selected
significand bit is only $b=27$.
As depth increases, the available range must accommodate
a longer chain, and the selected count falls to $b=10$
at $d=8$.
The errors of OzII-NoRescale consequently increase to
approximately $10^{-3}$ on all three datasets.
OzII-Rescale uses the same significand bits but also
introduces rounding between products.
Its depth-eight errors reach $0.423$, $1.32$, and $1.00$
on Uniform, Smooth, and Wide160, respectively.
Although dropping moduli reduces the integer magnitude,
it also reduces the representation range.
These results show that rescaling without base extension
does not maintain high accuracy for the tested chains
on the fixed 15-modulus basis.

OzII-Composed maintains normwise errors below
$8\times10^{-16}$ across all tested depths and datasets.
OzII-RescaleBE also achieves small errors through $d=7$,
while retaining the intermediate tensor in residue form.
Its errors range from $2.51\times10^{-16}$ to
$5.66\times10^{-16}$ on Uniform.
They are larger on Smooth and Wide160, reaching
$1.69\times10^{-15}$ and $2.60\times10^{-15}$,
respectively.

Unlike OzII-Composed, OzII-RescaleBE rounds the represented
integers during intermediate rescaling.
Entries that are small relative to the rescaling divisor
are more sensitive to this rounding.
Cancellation in Smooth and the wide input magnitude range
in Wide160 can make such effects more pronounced.
The observed differences are consistent with this
sensitivity, although the tests do not separately measure
the contributions of encoding and rescaling errors.

At $d=8$, the normwise errors of OzII-RescaleBE rise to
$6.74\times10^{-14}$, $3.37\times10^{-13}$, and
$6.00\times10^{-13}$ on Uniform, Smooth, and Wide160.
This increase coincides with the additional modulus drop
after the seventh mode product, which introduces another
rounded division.
Base extension restores the representation range but
does not reverse this rounding.
Retaining $b=53$ at encoding therefore does not guarantee
the same accuracy at every chain depth.

GEMMul8-Composed remains below $8\times10^{-16}$ in
normwise error across all tests.
GEMMul8-KRON generally has larger errors on Uniform
and Smooth, although its errors remain below
$2\times10^{-15}$ across all three datasets.
Both configurations have a nominal significand bit
of $b=53$ at $n=8$.
A possible explanation is the scaling used when the
dense operator is cached through \texttt{skip\_scalA}.
Reusing the operator fixes its encoding scales, and
the cached execution path may select more conservative
scales for the tensor operand, increasing its encoding
error.

\ElementwiseAccuracyTable

Table~\ref{tab:accuracy-elementwise} reports maximum and
median elementwise errors at depths 3 and 8.
OzII-RescaleBE maintains substantially smaller median errors
than OzII-NoRescale and OzII-Rescale on all three datasets.
At $d=3$, its median errors range from
$1.87\times10^{-16}$ to $1.54\times10^{-15}$,
compared with approximately $10^{-9}$ to $10^{-8}$
for the two other persistent variants.
For Uniform, its median error is comparable to those
of FP64 einsum and the composed implementations.
The larger medians on Smooth and Wide160 indicate
greater sensitivity to the input distribution.

At $d=8$, the median errors of OzII-RescaleBE increase
to $1.56\times10^{-13}$, $2.00\times10^{-12}$, and
$2.27\times10^{-12}$ on Uniform, Smooth, and Wide160,
respectively.
Although larger than those of FP64 einsum and the
composed implementations, these errors remain many
orders of magnitude below those of OzII-NoRescale
and OzII-Rescale.
The former has median errors between
$1.48\times10^{-3}$ and $6.18\times10^{-3}$,
while the latter has a median of $1.00$ on every dataset.
Thus, base extension substantially improves the accuracy
of persistent execution on the fixed residue basis,
although rescaling still limits accuracy as the chain
becomes deeper.

Maximum elementwise errors are substantially larger
than the medians for every method.
For example, on Smooth at $d=3$, OzII-RescaleBE has
a maximum error of $4.52\times10^{-9}$ and a median
of $1.47\times10^{-15}$.
FP64 einsum also shows a large separation, with
corresponding errors of $9.35\times10^{-11}$ and
$4.33\times10^{-16}$.
These results are consistent with the sensitivity of
elementwise relative error to small reference values.
Consequently, high normwise accuracy can coexist with
much larger relative errors in individual output entries.

\subsection{Performance Evaluation}
% \subsubsection{Measurement Procedure}

Execution times are measured using CUDA events over multiple consecutive evaluations of the complete chain
to reduce the effect of timing noise.
The total elapsed time is divided by the number of repetitions to obtain the mean execution time.
The OzII implementations and FP64 einsum use CUDA events through PyTorch, while GEMMul8 provides an internal
CUDA-event timer for its library operations.
Device memory usage is measured through PyTorch's interface to \texttt{cudaMemGetInfo}.

\subsubsection{Throughput}
\label{sec:performance-equal-entries}

The throughput tests keep the total number of degrees of
freedom (DoFs) approximately constant while varying the
tensor mode size.
Each scalar tensor entry counts as one element-local DoF.
The tests use Uniform inputs, chain depth $d=3$, and
$n\in\{8,12,16,32,64,128,256\}$.
The batch size is $\lfloor2^{24}/n^3\rfloor$, giving
32768, 9709, 4096, 512, 64, 8, and 1 tensors, respectively.
Each evaluation of the complete chain therefore processes
$2^{24}$ DoFs, except at $n=12$, where it processes
$2^{24}-64$.
Throughput is calculated by dividing this DoF count by
the elapsed time for the complete chain, including tensor
encoding and reconstruction.
Operator preparation and host--device transfers are excluded.

\begin{figure*}[t]
  \centering
  \includegraphics[width=0.8\linewidth]{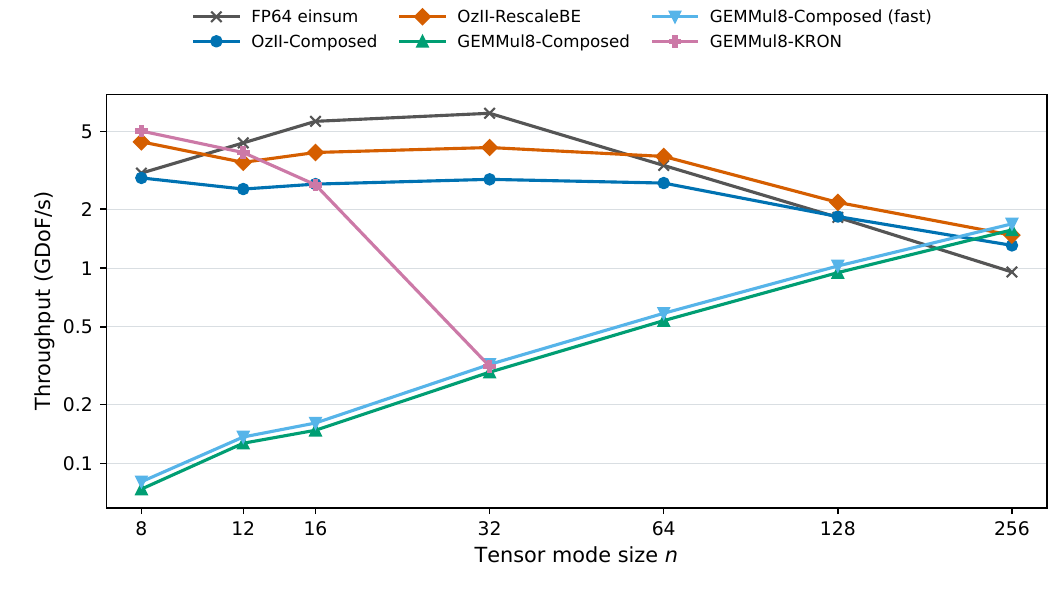}
  \caption{Throughput for three successive mode products
  on Uniform inputs with approximately $2^{24}$
  element-local DoFs per application.}
  \label{fig:performance-equal-entries}
\end{figure*}

Figure~\ref{fig:performance-equal-entries} shows the throughput of all compared methods.
FP64 einsum serves as the native-precision baseline.
Its throughput increases with tensor mode size up to
$n=32$ because the smaller mode products provide
insufficient work to keep the GPU busy.
At these sizes, execution is dominated by launch and
latency costs and by the permutation copies introduced
by einsum. As $n$ increases, cuBLAS processes larger, more efficient
GEMMs. From $n=64$, FP64 execution approaches the GPU's
double-precision throughput limit, so throughput
decreases approximately in proportion to the floating-point
operation count.

OzII-Composed has throughput similar to that of the
FP64 baseline at $n=8$, but falls below it for
$n=12$ through $64$.
Its three encodings and three reconstructions impose
a substantial conversion cost per DoF, while the FP64
DGEMMs become more efficient as $n$ grows.
The two methods have similar throughput at $n=128$.
At $n=256$, OzII-Composed achieves $1.37\times$ the
throughput of FP64 einsum.
This advantage is consistent with the increasing
importance of arithmetic throughput, because the
residue GEMMs use INT8 Tensor Cores rather than
native FP64 arithmetic.

At $n=8$, OzII-RescaleBE is $1.44\times$ higher than the FP64 baseline, whose small mode products incur
substantial launch and latency costs.
It is also $1.53\times$ higher than OzII-Composed
because it performs one encoding and one reconstruction
instead of three.
From $n=12$ to $n=32$, its throughput remains between
3.5 and 4.1\,GDoF/s, while the FP64 DGEMMs become
more efficient and reach their highest throughput
at $n=32$.
The FP64 baseline is therefore $1.2$--$1.5\times$ higher over these sizes.
From $n=64$, OzII-RescaleBE is $1.11$--$1.55\times$ higher than FP64 einsum as the latter becomes
increasingly constrained by FP64 arithmetic throughput.
Across all tested tensor mode sizes, OzII-RescaleBE is $1.13$--$1.53\times$ higher than OzII-Composed, 
showing the benefit of removing intermediate encodings and reconstructions.

GEMMul8-KRON outperforms the FP64 baseline at $n=8$
and OzII-RescaleBE at $n=8$ and $12$.
It applies the complete chain through a single
emulated GEMM with inner dimension $n^3$, providing
a longer contraction that is better suited to the
library's GEMM implementation.
Its throughput then falls rapidly because the dense
operator requires $n^3$ multiply--adds per output DoF,
compared with $3n$ for the three mode products.
The operator's $O(n^6)$ storage requirement also
limits its use at larger mode sizes.
At $n=32$, an FP64 copy of the operator occupies
approximately 8.6\,GB.
At $n=64$, it would require approximately 550\,GB,
exceeding the GPU's 96\,GB memory.
Therefore the measurements ends up at $n=32$.

The throughput of GEMMul8-Composed increases with
mode size over the tested range.
Larger $n$ increases the operator dimensions and
contracted length, making the mode products better
suited to the library's dense GEMM implementation.
Nevertheless, its throughput remains substantially
below that of the FP64 baseline and the OzII
implementations through $n=128$.
One reason is that the tested GEMMul8 configuration
pads internal buffer dimensions to multiples of 256.
At $n=8$, padding the operator row count and contracted
dimension considerably increases storage and data
movement.
At $n=256$, these dimensions require no padding,
and GEMMul8-Composed achieves $1.07\times$ the throughput
of OzII-RescaleBE, $1.21\times$ that of OzII-Composed,
and $1.65\times$ that of the FP64 baseline.
GEMMul8-Composed (fast) follows the same trend and
achieves $1.07$--$1.10\times$ the throughput of the
standard configuration at every tested size.

OzII-RescaleBE outperforms both GEMMul8-Composed
configurations through $n=128$, achieving
$55$--$60\times$ their throughput at $n=8$ and
$2.1$--$2.3\times$ at $n=128$.
Its throughput is 6--12\,\% lower at $n=256$.
It is also competitive with GEMMul8-KRON at small
tensor mode sizes and supports sizes beyond $n=32$,
where the dense operator becomes impractical.
Compared with the FP64 baseline, OzII-RescaleBE
achieves higher throughput at $n=8$ and from
$n=64$ onward.

\subsubsection{Execution-time Breakdown}
\label{sec:execution-time-breakdown}

\begin{figure*}[t]
  \centering
  \includegraphics[width=0.8\linewidth]{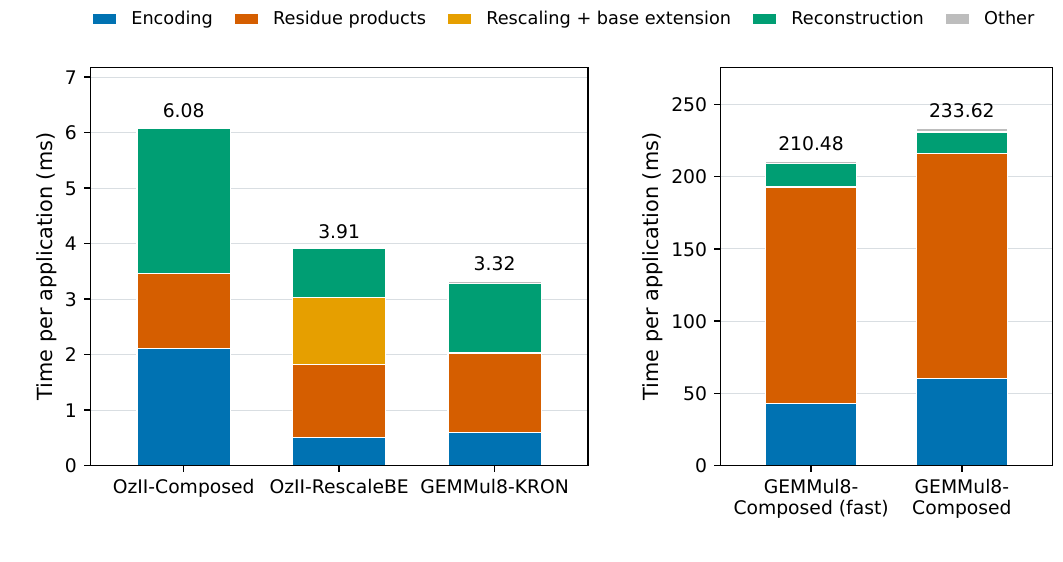}
  \caption{Execution-time breakdown at $n=8$, $d=3$,
  and $s=15$ for a batch of tensors containing a total
  of $2^{24}$ element-local DoFs.}
  \label{fig:breakdown-n8}
\end{figure*}

To better understand the preceding throughput results,
execution-time breakdowns are examined at $n=8$ and $256$.
Figure~\ref{fig:breakdown-n8} shows the breakdown for the
five compared methods at $n=8$.

The measurements are grouped into five categories.
Encoding includes scale discovery and conversion to residues.
Residue products include matrix multiplication and modular
reduction, which are fused in the OzII product kernels.
For GEMMul8, these operations are measured separately
but combined in the figures.
Rescaling and base extension form one category.
Reconstruction includes recovery from residues and inverse
scaling to FP64.
For GEMMul8, Other includes unfolding copies and the
difference between the measured library-call duration
and the sum of its internally measured phases.
Cached operator preparation is excluded.

For OzII-Composed, encoding and reconstruction dominate,
accounting for 4.73\,ms of the 6.08\,ms total and leaving
approximately 22\% for the residue products.
OzII-RescaleBE removes the intermediate encodings and
reconstructions, reducing their combined cost from
4.73 to 1.38\,ms.
However, rescaling and base extension add 1.21\,ms.
Together with the residue products, these operations
account for approximately 65\% of its execution time.
GEMMul8-KRON also encodes and reconstructs only once,
with these two steps taking 1.85\,ms in the dense
formulation.
Because it requires no rescaling or base extension,
its total execution time is 0.59\,ms shorter than
that of OzII-RescaleBE.

The two GEMMul8-Composed configurations are placed
in a separate panel with its own scale because their
execution times are much larger.
Residue products dominate the standard configuration,
accounting for approximately 156\,ms of its
233.62\,ms total.
Encoding is the second-largest cost at approximately
60\,ms, which the fast scaling path reduces to 43\,ms.

\begin{figure*}[t]
  \centering
  \includegraphics[width=0.8\linewidth]{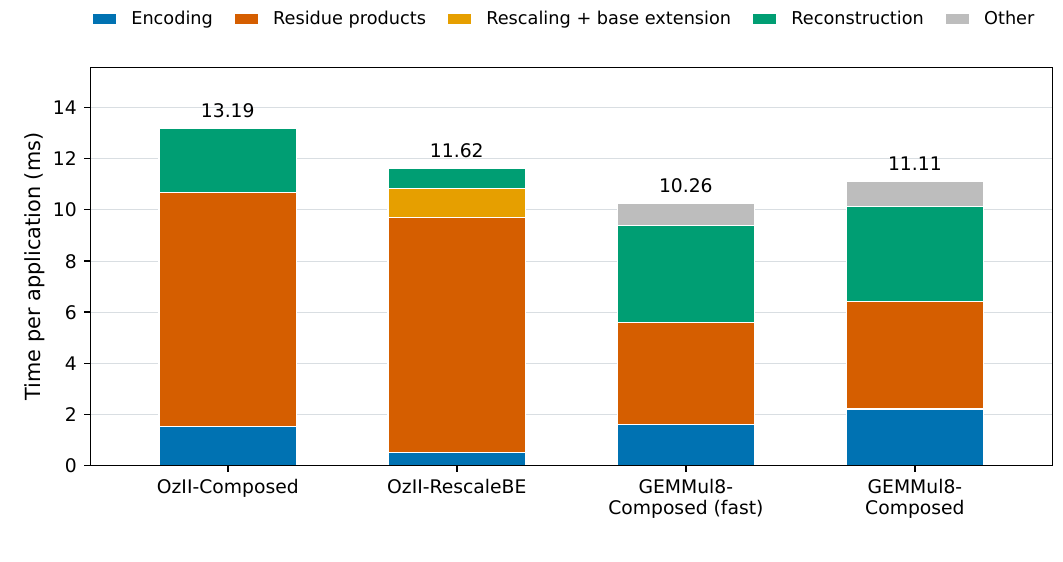}
  \caption{Execution-time breakdown at $n=256$, $d=3$,
  and $s=15$ for one tensor containing $2^{24}$
  element-local DoFs.}
  \label{fig:breakdown-n256}
\end{figure*}

Figure~\ref{fig:breakdown-n256} compares the four
remaining methods at $n=256$.
Residue products dominate both OzII implementations,
accounting for approximately 9.2\,ms of the 13.2\,ms
total for OzII-Composed and 9.2\,ms of the 11.6\,ms
total for OzII-RescaleBE.
OzII-Composed still spends approximately 31\%
(4.05\,ms) of its execution time on three encodings
and three reconstructions.
In contrast, OzII-RescaleBE spends only 21\% on
encoding, reconstruction, rescaling, and base extension
combined.

For the two GEMMul8-Composed configurations, the
distribution is reversed.
Encoding and reconstruction now dominate, accounting
for 5.4--5.9\,ms, or approximately 52--53\% of the
total execution time.
Although encoding and reconstruction are substantially
less costly in OzII-RescaleBE, the custom residue GEMM
kernels in the current implementation become the
bottleneck at large $n$.
This suggests that, for large contracted dimensions,
using cuBLAS INT8 GEMMs followed by a separate modular
reduction pass, as GEMMul8 does, could improve performance.

\subsubsection{Memory Usage}
\label{sec:memory-usage}

\begin{figure*}[t]
  \centering
  \includegraphics[width=0.5\linewidth]{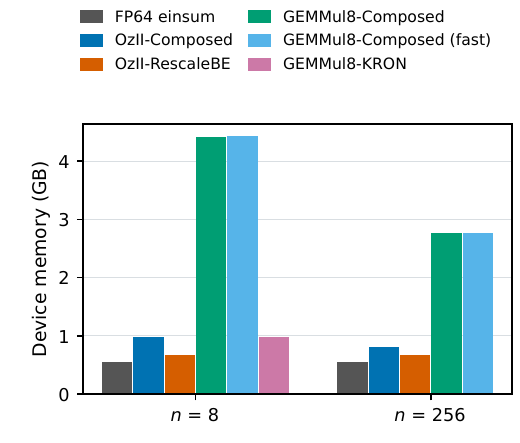}
  \caption{Device-memory footprint for tensor mode products
  on $2^{24}$ element-local DoFs with $d=3$,
  $n\in\{8,256\}$, and $s=15$.}
  \label{fig:memory-usage}
\end{figure*}

Figure~\ref{fig:memory-usage} reports the device-memory
footprint, measured as the increase in used device memory
from before input allocation to after setup, one warm-up
execution, and synchronization.
It includes the input and output tensors
(0.268\,GB together), the method's retained buffers,
and cached allocations and workspaces, but does not
capture transient peaks during execution.

At $n=8$, FP64 einsum occupies 0.539\,GB, including
approximately 0.27\,GB for two FP64 intermediates retained
by PyTorch's caching allocator.
OzII-Composed occupies 0.975\,GB, while OzII-RescaleBE
occupies 0.656\,GB, a reduction of 32.7\,\%.
GEMMul8-KRON occupies 0.965\,GB, and the two
GEMMul8-Composed configurations occupy 4.40 and 4.42\,GB.

The GEMMul8-Composed footprint at $n=8$ results from
the library's padding and workspace allocation.
Each mode product is a GEMM with dimensions $M=K=n$
and a column count equal to the batch size times $n^2$.
At $n=8$, this gives $M=K=8$ and 2,097,152 columns.
GEMMul8 rejects GEMMs with $2^{18}$ or more columns,
so the implementation splits the unfolding into
16 chunks of 131,072 columns, corresponding to
2,048 tensors each, and calls the library once per chunk.
The library pads $M$ and $K$ from 8 to 256.
For a fixed chunk, a buffer that scales linearly
with one of these padded dimensions is therefore
32 times larger than its unpadded counterpart.
One workspace per mode product is retained to preserve
the cached operator scaling.
The library reports 1.18\,GB of workspace per mode
product at $n=8$, compared with 0.62\,GB at $n=256$.

At $n=256$, the footprint of FP64 einsum is unchanged.
OzII-Composed occupies 0.797\,GB, while OzII-RescaleBE
occupies 0.656\,GB, a reduction of 17.7\,\%.
Both GEMMul8-Composed configurations occupy 2.743\,GB,
which is $4.18\times$ the footprint of OzII-RescaleBE.

The footprints of OzII-Composed and GEMMul8-Composed
decrease with increasing mode size for different reasons.
OzII-Composed scales each contracted fiber separately,
so its scale arrays and scale-discovery temporaries
grow in proportion to the number of fibers.
At $n=8$, there are approximately two million fibers
across 32,768 tensors, each containing 64 fibers,
requiring about 17\,MB per FP64 scale array.
At $n=256$, there are only 65,536 fibers,
requiring approximately 0.5\,MB per array.
OzII-RescaleBE uses one scale per tensor, and its
measured footprint remains unchanged.
For GEMMul8-Composed, the footprint decreases because
the GEMM dimensions at $n=256$ are already multiples
of 256 and require no padding.

\section{Conclusion}
\label{sec:conclusion}

This work presented OzII-RescaleBE, a persistent execution
method for applying Ozaki scheme~II to chained tensor
mode products.
The method combines rescaling and base extension to
control integer magnitude and restore representation
capacity between products.
It preserves the tensor mode-product structure while
encoding the input once, retaining intermediate tensors
in residue form, and reconstructing only the final result.
Successive mode products can therefore be evaluated without
converting intermediate tensors back to FP64.
An implementation using CuTe DSL and INT8 Tensor Cores
was provided for verification and performance evaluation.

Across three datasets with different input distributions,
OzII-RescaleBE achieved normwise relative errors ranging
from $2.5\times10^{-16}$ to $2.6\times10^{-15}$ for
chain depths $d=3$ through $7$.
The corresponding errors ranged from $10^{-16}$ to
$10^{-15}$ for GEMMul8-KRON and from $1\times10^{-16}$
to $3\times10^{-16}$ for GEMMul8-Composed and the
FP64 chain.
At $d=8$, the errors of OzII-RescaleBE increased to
between $7\times10^{-14}$ and $6\times10^{-13}$.

OzII-RescaleBE achieved throughputs ranging from
1.47 to 4.42\,GDoF/s across mode sizes $n=8$ to $256$.
Its throughput was within 12\,\% of GEMMul8-KRON at
$n=8$ and $12$ and higher at the larger sizes where
both methods were evaluated.
OzII-RescaleBE supported every tested size, whereas
GEMMul8-KRON was limited to $n\leq32$ by its
$O(n^6)$ operator storage requirement.
OzII-RescaleBE also achieved $2.1$--$60\times$ the
throughput of the GEMMul8-Composed configurations
through $n=128$.
The largest gains reflect the library's padding
overhead for small contracted dimensions.
At $n=256$, its throughput was 6--12\,\% lower
because the custom residue GEMM kernels were slower
than the library's cuBLAS-based implementation.
Compared with the FP64 baseline, OzII-RescaleBE
achieved higher throughput at $n=8$ and from
$n=64$ onward.

The execution-time breakdown explains these differences.
At $n=8$, rescaling and base extension introduced
additional costs that outweighed the conversion savings
relative to GEMMul8-KRON.
At $n=256$, residue products accounted for approximately
79\,\% of the OzII-RescaleBE execution time.
Its device-memory footprint at $n=8$ was approximately
32\,\% smaller than that of GEMMul8-KRON, while
GEMMul8-Composed required approximately $6.7\times$
as much memory.
At $n=256$, GEMMul8-Composed required approximately
$4.2\times$ as much memory, with its retained workspaces
for individual mode products contributing to the
difference even without padding.

Beyond the performance limitations of the current
implementation, the method has two main limitations.
First, rescaling introduces rounding errors, as reflected
in the increased errors at depth eight.
A larger modulus basis may permit less aggressive
rescaling, but also increases storage and residue-product
work.
Second, the current formulation uses a single encoding
scale for each input tensor.
When tensor entries span a wide range of magnitudes,
small entries can lose significant bits or round to
zero during the initial encoding.
Unlike per-fiber encoding in OzII-Composed, this shared
scale cannot adapt to the magnitude distribution of
individual fibers.
Any resulting information loss occurs before rescaling
and cannot be recovered by base extension.
% Second, persistent execution does not eliminate the
% cost of processing intermediate tensors.
% It replaces intermediate floating-point conversions
% with rescaling and base extension, whose costs were
% lower in the tested comparisons with OzII-Composed
% but remain part of the execution time.

Future work includes improving the residue-product
kernels and exploring shared-memory execution that
retains intermediate residues on chip for small tensors.
A native CUDA implementation will support integration
into DG simulations, followed by evaluation of
representative operators over a broader range of
tensor sizes, chain depths, and modulus counts.
Alternatives to greedy modulus dropping will also
be investigated to account for later mode products
and balance rounding error, representation capacity,
and execution cost.

\section*{Acknowledgements}
This work is partially supported by MEXT as "Project for Establishment of a Center for 
Advanced HPC-AI Development Support".

Generative AI tools were used under the authors' direction. Claude Code (Fable~5.1)
assisted with the implementation, the evaluation scripts and measurements, the figures,
and the drafting of parts of the text. GPT-6 Astra generated initial manuscript drafts
and assisted with revision. The authors directed the investigation, verified the
AI-generated material against exact references and independent measurements, and take
responsibility for the scientific content.

\clearpage
\bibliographystyle{unsrt}  
\bibliography{references}

@article{swirydowicz2019tensor,
  author = {{\'{S}}wirydowicz, Kasia and Chalmers, Noel and Karakus, Ali and Warburton, Tim},
  title = {Acceleration of tensor-product operations for high-order finite element methods},
  journal = {The International Journal of High Performance Computing Applications},
  year = {2019},
  volume = {33},
  number = {4},
  pages = {735--757},
  doi = {10.1177/1094342018816368},
  url = {https://arxiv.org/abs/1711.00903}
}

@article{ozaki2012eft,
  author = {Ozaki, Katsuhisa and Ogita, Takeshi and Oishi, Shin'ichi and Rump, Siegfried M.},
  title = {Error-free transformations of matrix multiplication by using fast routines of matrix multiplication and its applications},
  journal = {Numerical Algorithms},
  year = {2012},
  volume = {59},
  number = {1},
  pages = {95--118},
  doi = {10.1007/s11075-011-9478-1},
  url = {https://doi.org/10.1007/s11075-011-9478-1}
}

@inproceedings{mukunoki2020tensor,
  author = {Mukunoki, Daichi and Ozaki, Katsuhisa and Ogita, Takeshi and Imamura, Toshiyuki},
  title = {{DGEMM} Using Tensor Cores, and Its Accurate and Reproducible Versions},
  booktitle = {High Performance Computing},
  year = {2020},
  publisher = {Springer},
  pages = {230--248},
  doi = {10.1007/978-3-030-50743-5\_12},
  url = {https://doi.org/10.1007/978-3-030-50743-5\_12}
}

@article{ootomo2024integer,
  author = {Ootomo, Hiroyuki and Ozaki, Katsuhisa and Yokota, Rio},
  title = {{DGEMM} on Integer Matrix Multiplication Unit},
  journal = {The International Journal of High Performance Computing Applications},
  year = {2024},
  doi = {10.1177/10943420241239588},
  url = {https://arxiv.org/abs/2306.11975}
}

@article{ozaki2025scheme2,
  author = {Ozaki, Katsuhisa and Uchino, Yuki and Imamura, Toshiyuki},
  title = {{Ozaki Scheme II}: A {GEMM}-oriented emulation of floating-point matrix multiplication using an integer modular technique},
  journal = {The International Journal of High Performance Computing Applications},
  year = {2026},
  doi = {10.1177/10943420261467787},
  url = {https://arxiv.org/abs/2504.08009v3}
}

@misc{gemmul8,
  author = {{RIKEN R-CCS}},
  title = {{GEMMul8}: {GEMM} emulation using low-precision matrix engines based on {Ozaki Scheme II}},
  howpublished = {Software repository},
  url = {https://github.com/RIKEN-RCCS/GEMMul8},
  note = {Accessed 2026-09-11}
}

@misc{lu2026emugemm,
  author = {Lu, Denghui and Maeder, Alexander and Luisier, Mathieu and Ziogas, Alexandros Nikolaos},
  title = {{EmuGEMM}: Fused Tensor Core Kernels for Precision Emulation in Matrix Multiplication},
  year = {2026},
  archivePrefix = {arXiv},
  eprint = {2606.25453},
  primaryClass = {cs.DC},
  url = {https://arxiv.org/abs/2606.25453v1}
}

@inproceedings{cheon2017ckks,
  author = {Cheon, Jung Hee and Kim, Andrey and Kim, Miran and Song, Yongsoo},
  title = {Homomorphic Encryption for Arithmetic of Approximate Numbers},
  booktitle = {Advances in Cryptology -- ASIACRYPT 2017},
  year = {2017},
  series = {Lecture Notes in Computer Science},
  volume = {10624},
  publisher = {Springer},
  doi = {10.1007/978-3-319-70694-8\_15},
  url = {https://eprint.iacr.org/2016/421}
}

@misc{cheon2018rns,
  author = {Cheon, Jung Hee and Han, Kyoohyung and Kim, Andrey and Kim, Miran and Song, Yongsoo},
  title = {A Full {RNS} Variant of Approximate Homomorphic Encryption},
  howpublished = {Cryptology ePrint Archive, Paper 2018/931},
  year = {2018},
  url = {https://eprint.iacr.org/2018/931},
  note = {The ePrint record identifies SAC 2018 publication}
}

@misc{halevi2018rns,
  author = {Halevi, Shai and Polyakov, Yuriy and Shoup, Victor},
  title = {An Improved {RNS} Variant of the {BFV} Homomorphic Encryption Scheme},
  howpublished = {Cryptology ePrint Archive, Paper 2018/117},
  year = {2018},
  url = {https://eprint.iacr.org/2018/117},
  note = {Revised 2019-03-06; published at CT-RSA 2019, DOI 10.1007/978-3-030-12612-4\_5}
}

@misc{nvidiaH100,
  author = {{NVIDIA}},
  title = {{NVIDIA H100 GPU}: Product Specifications},
  howpublished = {Manufacturer product page},
  url = {https://www.nvidia.com/en-us/data-center/h100/},
  note = {H100 SXM specifications; accessed 2026-09-11}
}

@misc{nvidiaBlackwellUltra,
  author = {{NVIDIA}},
  title = {{NVIDIA Blackwell Ultra}: Datasheet},
  howpublished = {Manufacturer datasheet},
  url = {https://dam-cdn.nvd.orangelogic.com/AssetLink/1k0p832eq8r5ca0u5383ie5o4tp3bst1.pdf},
  note = {Technical specifications, p. 5; accessed 2026-09-11}
}

@article{kolda2009tensor,
  author = {Kolda, Tamara G. and Bader, Brett W.},
  title = {Tensor Decompositions and Applications},
  journal = {SIAM Review},
  year = {2009},
  volume = {51},
  number = {3},
  pages = {455--500},
  doi = {10.1137/07070111X},
  url = {https://www.kolda.net/publication/TensorReview.pdf}
}

@misc{schwarz2025guaranteed,
  author = {Schwarz, Angelika and Anders, Anton and Brower, Cole and Bayraktar, Harun and Gunnels, John and Clark, Kate and Xu, RuQing G. and Rodriguez, Samuel and Cayrols, Sebastien and Tabaszewski, Pawe{\l} and Podlozhnyuk, Victor},
  title = {Guaranteed {DGEMM} Accuracy While Using Reduced Precision Tensor Cores Through Extensions of the {Ozaki} Scheme},
  year = {2025},
  archivePrefix = {arXiv},
  eprint = {2511.13778},
  primaryClass = {cs.DC},
  url = {https://arxiv.org/abs/2511.13778},
  note = {Also published at SCA/HPCAsia 2026, DOI 10.1145/3773656.3773670}
}

@misc{hayashi2026fp4,
  author = {Hayashi, Shun-ichiro and Mukunoki, Daichi and Hoshino, Tetsuya and Katagiri, Takahiro},
  title = {{DGEMM} with {Ozaki Scheme I/II} on {FP4} Tensor Cores: A Base-13 {E2M1} Limb Representation},
  year = {2026},
  archivePrefix = {arXiv},
  eprint = {2608.06812},
  url = {https://arxiv.org/html/2608.06812}
}

@inproceedings{mirage2023,
  author = {Demirkiran, Cansu and Yang, Guowei and Bunandar, Darius and Joshi, Ajay},
  title = {Mirage: An {RNS}-Based Photonic Accelerator for {DNN} Training},
  booktitle = {2024 ACM/IEEE 51st Annual International Symposium on Computer Architecture (ISCA)},
  year = {2024},
  doi = {10.1109/ISCA59077.2024.00016},
  url = {https://arxiv.org/abs/2311.17323v2}
}

@misc{uchino2026fp8,
  author = {Uchino, Yuki and Ozaki, Katsuhisa and Imamura, Toshiyuki},
  title = {Double-Precision Matrix Multiplication Emulation via {Ozaki-II} Scheme with {FP8} Quantization},
  year = {2026},
  archivePrefix = {arXiv},
  eprint = {2603.10634},
  primaryClass = {cs.DC},
  url = {https://arxiv.org/abs/2603.10634v2}
}

@misc{bajard2016rns,
  author = {Bajard, Jean-Claude and Eynard, Julien and Hasan, Anwar and Zucca, Vincent},
  title = {A Full {RNS} Variant of {FV} like Somewhat Homomorphic Encryption Schemes},
  howpublished = {Cryptology ePrint Archive, Paper 2016/510},
  year = {2016},
  url = {https://eprint.iacr.org/2016/510},
  note = {Extended version of the SAC 2016 paper}
}

@misc{mukunoki2025fp8,
  author = {Mukunoki, Daichi},
  title = {{DGEMM} without {FP64} Arithmetic -- Using {FP64} Emulation and {FP8} Tensor Cores with {Ozaki} Scheme},
  year = {2025},
  archivePrefix = {arXiv},
  eprint = {2508.00441},
  url = {https://arxiv.org/abs/2508.00441v3}
}

@misc{uchino2025complex,
  author = {Uchino, Yuki and Ma, Qianxiang and Imamura, Toshiyuki and Ozaki, Katsuhisa and Gutsche, Patrick Lars},
  title = {Emulation of Complex Matrix Multiplication based on the {Chinese Remainder Theorem}},
  year = {2025},
  archivePrefix = {arXiv},
  eprint = {2512.08321},
  primaryClass = {cs.DC},
  url = {https://arxiv.org/abs/2512.08321v3},
  note = {Inspected preprint version 3. The arXiv record also identifies ISC High Performance 2026 proceedings, DOI 10.23919/isc.2026.11520500}
}

@misc{matsuoka2026fft,
  author = {Matsuoka, Satoshi},
  title = {{FP8} is All You Need (Part 2): Full-{FP64} 3-D {FFT} on {FP8}-Generation Tensor Cores: The Integer-Epilogue Wall and the Minimal Hardware That Would Remove It},
  year = {2026},
  archivePrefix = {arXiv},
  eprint = {2606.23698},
  url = {https://arxiv.org/abs/2606.23698v3},
  note = {Version 3, 8 September 2026}
}

@article{garner1959residue,
  author = {Garner, H. L.},
  title = {The Residue Number System},
  journal = {IRE Transactions on Electronic Computers},
  year = {1959},
  volume = {EC-8},
  number = {2},
  pages = {140--147},
  month = jun,
  doi = {10.1109/TEC.1959.5219515},
  url = {https://doi.org/10.1109/TEC.1959.5219515}
}

@article{dumas2008fflas,
  author = {Dumas, Jean-Guillaume and Giorgi, Pascal and Pernet, Cl{\'e}ment},
  title = {Dense Linear Algebra over Word-Size Prime Fields: The {FFLAS} and {FFPACK} Packages},
  journal = {ACM Transactions on Mathematical Software},
  year = {2008},
  volume = {35},
  number = {3},
  pages = {1--42},
  doi = {10.1145/1391989.1391992},
  url = {https://doi.org/10.1145/1391989.1391992}
}

@misc{nvidiaCute452Warp,
  author = {{NVIDIA Corporation}},
  title = {{CuTe DSL} 4.5.2 Warp Submodule},
  year = {2026},
  url = {https://docs.nvidia.com/cutlass/4.5.2/media/docs/pythonDSL/cute_dsl_api/cute_nvgpu_warp.html},
  note = {Accessed September 14, 2026}
}

@misc{nvidiaRtxProBlackwellArchitecture,
  author = {{NVIDIA}},
  title = {{NVIDIA RTX PRO Blackwell GPU Architecture}},
  howpublished = {Architecture white paper, version 1.1},
  url = {https://www.nvidia.com/content/dam/en-zz/Solutions/design-visualization/quadro-product-literature/pdf/NVIDIA-RTX-Blackwell-PRO-GPU-Architecture-v1_1.pdf},
  note = {FP64 throughput ratio on p. 8 and Max-Q specifications in Table 4 on pp. 46--47. Accessed 2026-09-15}
}

@misc{nvidiaPtxMma,
  author = {{NVIDIA Corporation}},
  title = {{PTX ISA}, Warp-Level Matrix Multiply-Accumulate Instructions},
  url = {https://docs.nvidia.com/cuda/parallel-thread-execution/index.html#warp-level-matrix-instructions-mma},
  note = {Accessed September 16, 2026}
}

@article{cui2024tensor,
  author  = {Cui, Cu},
  title   = {Acceleration of Tensor-Product Operations with Tensor Cores},
  journal = {ACM Transactions on Parallel Computing},
  year    = {2024},
  volume  = {11},
  number  = {4},
  doi     = {10.1145/3695466}
}

@misc{tu2026fp64,
  author        = {Tu, Jiqun and Karlin, Ian and Camier, John
                   and Dobrev, Veselin and Kolev, Tzanio
                   and Henneking, Stefan and Ghattas, Omar},
  title         = {Accelerating High-Order Finite Element Simulations
                   at Extreme Scale with {FP64} Tensor Cores},
  year          = {2026},
  eprint        = {2603.09038},
  archivePrefix = {arXiv},
  primaryClass  = {cs.DC},
  doi           = {10.48550/arXiv.2603.09038}
}

@article{horner1819method,
  author  = {Horner, W. G.},
  title   = {A New Method of Solving Numerical Equations of All Orders,
             by Continuous Approximation},
  journal = {Philosophical Transactions of the Royal Society of London},
  volume  = {109},
  pages   = {308--335},
  year    = {1819},
  doi     = {10.1098/rstl.1819.0023}
}

@article{higham2022mixed,
  author  = {Higham, Nicholas J. and Mary, Theo},
  title   = {Mixed precision algorithms in numerical linear algebra},
  journal = {Acta Numerica},
  volume  = {31},
  pages   = {347--414},
  year    = {2022},
  doi     = {10.1017/S0962492922000022}
}

\clearpage
\appendix
\section{Additional comparison methods}
\label{app:comparison-methods}

This appendix describes the two additional persistent
Ozaki-II variants and the FP64 einsum used in the evaluation.
OzII-NoRescale performs the chain without intermediate
rescaling.
OzII-Rescale drops moduli through rescaling but does not
restore them.
Both encode the input tensor once and decode only the
final result.
They follow the effective-operator preparation and reconstruction
conventions of the main persistent method.

\subsection{FP64 einsum baseline}
\label{app:fp64-einsum}

The FP64 einsum applies one \texttt{torch.einsum} contraction
per mode product to CUDA tensors stored in
\texttt{torch.float64}.
Each call applies the corresponding $n\times n$ operator
along the specified mode, preserving the batch dimension
and the other two tensor indices.
Its FP64 output becomes the input to the next mode product,
following the same operator order and mode sequence as
the residue implementations.
The baseline does not form a dense Kronecker operator or
perform residue encoding.
It provides a finite-precision execution baseline,
distinct from the exact rational-chain reference whose
final result is converted to FP64 for error measurement.

\subsection{OzII-NoRescale}
\label{app:OzII-NoRescale}

For a candidate significand bits $b$, let $\bar A_i(b)$
denote the effective integer operator for the $i$-th mode
product.
Its maximum absolute row sum is
\begin{equation}
  \rho_i(b)=\max_{1\leq j\leq n}
    \sum_{k=1}^{n}|(\bar A_i(b))_{jk}|.
  \label{eq:appendix-row-sum}
\end{equation}
Each output entry is a weighted sum along one contracted
fiber, so multiplying the current magnitude bound by
$\rho_i(b)$ gives a bound for the output entries.
The effective operators and their bounds are recomputed
for each candidate $b$.

The input uses one encoding scale per tensor, giving
the initial reconstruction scale $\alpha=2^{-e}$.
The field $\Gamma$ contains the remaining operator
reconstruction scales.
When a mode is revisited, its pending scales are folded
into the next effective operator before encoding,
as in the main methodology.
Thus, $\Gamma$ contains only the scales still to be
applied at final reconstruction.

OzII-NoRescale retains the initial basis $\mathcal P$
throughout the chain, with $P=\prod_{p\in\mathcal P}p$.
Without rescaling, the integer magnitude after the
$j$-th mode product is bounded by
\begin{equation}
  B_j(b)=2^b\prod_{i=1}^{j}\rho_i(b),
  \qquad B_0(b)=2^b.
  \label{eq:appendix-no-rescale-bound}
\end{equation}
The planner selects the largest $b\in\{1,\ldots,53\}$
satisfying $2B_j(b)<P$ for every $j=0,\ldots,d$.
When all $\rho_i(b)\geq1$, the bounds are nondecreasing,
and this condition reduces to the whole-chain test
$2^b\prod_{i=1}^{d}\rho_i(b)<P/2$.
The range requirement can reduce the selected encoding
bit count as depth increases, even though the number
of residue planes remains constant.
The method retains the supplied $s$ moduli and does
not require the entire available modulus pool.

\NoRescaleAppendixAlgorithm

\subsection{OzII-Rescale and its planner}
\label{app:OzII-Rescale}

OzII-Rescale starts from the same basis but removes moduli
after nonfinal mode products.
Algorithm~\ref{alg:app-planrescale} calls
\texttt{DropPlan} with restoration disabled.
The flag $\tau=0$ corresponds to the \textsc{Shrink}
mode in the implementation.
In contrast, $\tau=1$, used by \texttt{PlanRescaleBE},
assumes that base extension restores the working basis
before the next mode product.

The planner starts at $b=53$ and decreases $b$ until
the range checks pass for every product and planned drop.
It returns the selected significand bits and the
drop counts $(g_i)$, where $g_i$ specifies the number
of drops after the $i$-th mode product and $g_d=0$.
If no positive $b$ passes the checks, it reports failure.
No base extension is performed, so removed moduli
remain unavailable for subsequent products.

\PlanRescaleAppendixAlgorithm

Algorithm~\ref{alg:app-OzIIrescale} maintains the available
basis $\mathcal L$, initialized to $\mathcal P$.
At each planned drop, \texttt{RnsDrop} divides the
represented integer by the smallest available modulus
$q$ with rounding, removes that modulus, and updates
the reconstruction scale to $\alpha'=\alpha q$.
The returned scale becomes $\alpha$ for the next operation.
At the end of the chain, $\alpha$ therefore contains
the initial tensor reconstruction scale multiplied by every
dropped modulus.

The final reconstruction uses the remaining basis and the
scale $\alpha\Gamma$.
Rescaling reduces both the integer magnitude and the
available representation range.
Without base extension, that lost representation
capacity is not restored.

\RescaleAppendixAlgorithm

\end{document}